\documentclass[twocolumn]{aastex63}
\usepackage{amsmath,bm}

\submitjournal{ApJ}

\shorttitle{3D Flow Field of Eccentric Planets}
\shortauthors{Bailey et al.}
\graphicspath{{./}{figures/}}

\begin{document}
\title{The Three Dimensional Flow Field Around Planets on Eccentric Orbits}

\correspondingauthor{Avery Bailey}
\email{apbailey@princeton.edu}

\author{Avery Bailey}
\affiliation{Department of Astrophysical Sciences,
Princeton University, Princeton, NJ 08544, USA}

\author{James M. Stone}
\affiliation{School of Natural Sciences, Institute for Advanced Study, Princeton, NJ 08544, USA}
\affiliation{Department of Astrophysical Sciences,
Princeton University, Princeton, NJ 08544, USA}

\author{Jeffrey Fung}
\affiliation{School of Natural Sciences, Institute for Advanced Study, Princeton, NJ 08544, USA}

\nocollaboration{3}

\begin{abstract}
We investigate the properties of the hydrodynamic flow
around eccentric protoplanets and compare them with the often assumed 
case of a circular orbit. To this end, we perform 
a set of 3D
hydrodynamic simulations of protoplanets with small eccentricities 
($e\leq 0.1$). We adopt an isothermal equation of state and 
concentrate resolution on the protoplanet to investigate flows down 
to the scale of the protoplanet’s circumplanetary disk (CPD). 
We find enhanced prograde rotation exterior to the CPD for low planet 
masses undergoing subsonic eccentric motion. 
If the eccentricity is made large enough to develop a bow shock, 
this trend reverses and rotation becomes increasingly retrograde. 
The instantaneous eccentric flow 
field is dramatically altered compared to circular orbits. Whereas the latter
exhibit a generic pattern of polar inflow and 
midplane outflow, the flow geometry depends on orbital phase in 
the eccentric case. 
For even the modest eccentricities tested here, 
the dominant source of inflow can come from the midplane instead 
of the poles. 
We find that 
the amount of inflow and outflow increases
for higher $e$ and lower protoplanet masses, thereby 
recycling more gas through the planet's Bondi radius.
These increased fluxes may increase the 
pebble accretion rate for eccentric planets up to
several times that of the circular orbit rate.
In response to eccentric motion, the structure and rotation of the 
planet's bound CPD remains unchanged. 
Because the CPD regulates the eventual accretion of gas 
onto the planet, we predict little change to the gas accretion rates 
between eccentric and circular planets.
\end{abstract}

\keywords{editorials, notices --- 
miscellaneous --- catalogs --- surveys}

\section{Introduction} \label{sec:intro}
While classic 1D models \citep{Mizuno1980,Pollack+1996} 
model core accretion as 
the quasi-static contraction of a surrounding envelope,
recent simulations of protoplanets are painting an increasingly 
hydrodynamic and three-dimensional picture. 
Early 3D simulations \citep{Machida+2008,AyliffeBate2012,Tanigawa+2012} identified  
a robust pattern of polar inflow coupled with midplane outflow.
This meridional circulation is sourced by high latitude horseshoe 
flows that descend and penetrate deep into the atmosphere before 
outflowing in the midplane \citep{Fung+2015}. This circulation 
occurs for a wide range of planet masses -- from super-Earths 
\citep{Ormel+2015} and giant planet cores \citep{LambrechtsLega2017} 
to gas giants \citep{Schulik+2019}. 
The pattern continues to persist even with more sophisticated 
treatment of the thermodynamics \citep{DangeloBodenheimer2013}. 
It has been argued that for super-Earth sized protoplanets, gas 
recycling associated with meridional circulation is dominant enough 
to stave off the transition to runaway accretion 
\citep{Ormel+2015,Cimerman+2017}. 
Even in simulations where a bound atmosphere emerges 
\citep{LambrechtsLega2017,BethuneRafikov2019},
recycling can alter the region’s size, making it smaller than 
1D models would assume.
This suggests that the 3D hydrodynamic nature of the 
flow field should be taken into account in building a comprehensive theory of 
planet formation.

Though less robust than meridional circulation, a 
number of studies have also seen the emergence of a circumplanetary 
disk (CPD). This inherently multi-dimensional phenomenon tends 
to occur preferentially towards higher mass planets 
\citep{Machida2009,AyliffeBate2009,Tanigawa+2012}. 
More recent studies have shown that CPDs are also sensitive 
to the thermodynamics of the gas \citep{Szulagyi+2016,Szulagyi2017,Fung+2019}, 
occurring most readily in isothermal atmospheres \citep{Tanigawa+2012,Wang+2014,Fung+2019}. 
When and if a CPD forms, in turn changes 
the nature of accretion onto the protoplanet \citep{Schulik+2020} and relates to questions of moon formation. 

All of these simulations however, 
adopt the assumption of a circular orbit. In this study, 
we investigate how robust this and other results of the 3D 
simulations are when the circular orbit assumption is relaxed and 
replaced with a low to moderate eccentricity $(e\leq 0.1)$. 
Whether we are justified in allowing for eccentric protoplanets 
requires examining the balance of torques which either damp or 
excite eccentricity. Low mass embedded planets and planetesimals 
are expected to have their eccentricity damped 
through interactions with the surrounding disk. 
In linear theory, gravitational torques 
exerted at exterior Lindblad resonances serve to increase a planet's 
eccentricity, opposite signed torques exerted at co-orbital Lindblad 
resonances are large enough to overwhelm the exterior torques 
and make the net evolution one of 
eccentricity damping \citep{Artymowicz1993}. The damping time 
for planet with mass $M_p$ embedded in a disk with aspect ratio $h_p$ 
is \citep{Artymowicz1993}:
\begin{equation}
    t_e \approx 3\times10^3 \text{ yrs }\left(\frac{M_\oplus}{M_p}\right)
    \left(\frac{h_p}{0.05}\right)^4
\end{equation}
with additional scalings dependent on disk properties omitted and 
assuming $e\lesssim h_p$. For $e\gtrsim h_p$, 
\citet{PapaloizouLarwood2000} suggest an additional scaling 
$\left(e/h_p\right)^3$ making the circularization times longer, 
but still shorter than the formation timescale for planets. 
Torques from numerical simulations of embedded planets further suggest that eccentricity is damped \citep{BitschKley2010,FendykeNelson2014}.

For planets sufficiently massive to open a gap, the co-orbital 
Lindblad contributions vanish, removing the main source of 
eccentricity damping \citep{GoldreichSari2003}. 
At this point in the planet's evolution, the 
response is highly non-linear, requiring full hydrodynamic 
simulations to adequately compute the torques. 
Simulations suggest that 
planets 
can become eccentric in gaps 
\citep{Dangelo+2006,Dunhill+2013,DuffellChiang2015,Muley+2019}. 
Though these results are generally for higher mass planets than 
those investigated here, similar behavior could apply to lower masses given a sufficiently low viscosity.
 
Furthermore, the damping mechanisms described above assume
a fairly quiescent and passive formation process. In all likelihood, 
the formation pathways are far more complex, allowing for modes of 
exciting eccentricity. Towards low masses for example, a protoplanet 
swarm in equlibrium with its protoplanetary disk is expected to 
produce characteristic eccentricities on the order of $h_p$
\citep{PapaloizouLarwood2000}. 
A stellar companion or partner giant planet interacting through 
secular or mean-motion resonances could also excite these moderate 
eccentricities. Even heating effects can result in eccentricities 
close to $h_p$ \citep{EklundMasset2017,Chrenko+2017}. Though the 
relative importance of these effects is uncertain, it seems unlikely 
that the circular orbit assumption applies to all protoplanets.
In Section \ref{sec:damp}, with insight gained from our simulations,
we will further call into question the assumption of damped 
eccentricity.

Few works have explored the modifications that eccentric 
protoplanets would make to modern theories of planet formation. 
First there is the work of \citet{LiuOrmel2018,OrmelLiu2018} who 
perform 2D and 3D 
pebble integrations to understand how pebble accretion 
efficiencies might change given an accreting eccentric protoplanet 
instead of the often assumed circular one. Their work suggests an 
enhancement to the pebble accretion efficiency for low 
eccentricities $e\sim0.05$ with an eventual suppression as one 
goes to even higher eccentricity $e\sim0.1$. This does not 
however, take into account perturbations to the flow field 
arising from the planetary potential—perturbations which, 
at least in the circular case, have been shown to dramatically 
suppress the accretion of small pebbles. More recently, 
\citet{Mai+2020} who perform 2D radiation hydrodynamic simulations 
of planets on highly eccentric $e>0.1$ orbits. At such high $e$, 
the dynamics becomes dramatically different from circular models 
due to the emergence of a bow shock. 
What happens in the $e<0.1$ regime however? 
This regime is physically interesting because it spans the 
transition from subsonic to supersonic orbital motion. Is the 
flow field altered by the planet as in the circular case or is 
it determined by eccentric motion?

To answer these questions we perform global 3D isothermal 
hydrodynamical simulations of planets on circular and eccentric 
$(e\leq 0.1)$ orbits and study how the flow field is altered. 
Our simulations end up being entirely complementary to those by 
\citet{Mai+2020}. We sacrifice their more thorough treatment 
of thermodynamics for the ability to study the inherently 3D flows. 
We work in a global framework whereas they adopt the local 
approximation. We end our simulations with $e=0.1$, precisely where 
their models begin. In this way, we are able to begin to build a 
unified picture of eccentric protoplanets and test the relative 
importance of relevant physics. 

Because the work on eccentric protoplanets has been fairly sparse, 
this study is largely exploratory. We explore the most information 
dense regions of parameter space where the planet is both transitioning 
from subsonic to supersonic ($e\leq 0.1$) and transitioning 
from the linear to non-linear mass regime. This allows us to explore 
a maximal amount of interesting regimes with the fewest number of 
computationally expensive models. The drawback of course being that 
we are unable to explore any of these regimes in particular detail. 
We instead extrapolate based on the few models run in this transition 
regime. This strategy necessitates further confirmation of the 
extrapolated results by more focused studies but provides an 
exploratory glimpse of the how eccentricity can 
modify the 3D structure of the protoplanetary envelope.

We begin this paper with a description of our methods (Section \ref{sec:methods}). 
In Section \ref{sec:e0} we discuss the circular orbit models and 
demonstrate their consistency with previous works. 
Section \ref{sec:ecc} gives the results of our eccentric models and 
compares them with the established patterns of circular 
orbit models. Finally in Section \ref{sec:imp} we discuss potential 
implications of these eccentric models and questions of their 
applicability.

\section{Methods} \label{sec:methods}
\subsection{Equations and Code}\label{sec:eqns}
To determine how protoplanet eccentricity might alter current 
theories of planet formation, we perform a series of 3D 
global hydrodynamic simulations of a protoplanet orbiting 
its host star through a protoplanetary disk.
In a frame rotating with uniform angular velocity $\bm{\Omega_0}$, 
the motion of nebular gas at position $\bm{r}$
is governed by the following fluid equations:
\begin{equation}\label{eq:f1}
\frac{\partial \rho}{\partial t}
+ \bm{\nabla\cdot}\left(\rho \bm{v}\right) = 0
\end{equation}
\begin{multline}\label{eq:f2}
\frac{\partial \left(\rho \bm{v}\right)}{\partial t} 
+ \bm{\nabla\cdot}\left(\rho \bm{v\otimes v} + \bm{P}\right) = 
- \rho\bm{\nabla} \Phi
- 2\rho\bm{\Omega_0\times v}\\ - \rho\bm{\Omega_0\times\Omega_0\times r}
\end{multline}
where $\rho$, $\bm{v}$, $\bm{P}$, $\Phi$ are the standard fluid density,
velocity, diagonal pressure tensor, and gravitational potential 
respectively.
For the sake of simplicity and because protoplanets 
generically form circumplanetary disks (CPDs) under isothermal 
conditions \citep{Fung+2015}, we adopt an isothermal 
equation of state,
\begin{equation}\label{eq:f3}
    P = \rho c^2
\end{equation}
with sound speed $c$, to close the above equations. 
Because of this choice, these models are able to test whether 
CPDs are disrupted or altered by eccentric motion. 

To solve equations \eqref{eq:f1}-\eqref{eq:f3}, 
we employ the \textsc{Athena++} 
code\footnote{\url{http://princetonuniversity.github.io/athena/}}
\citep{Stone+2020}.
Because \textsc{Athena++} 
does not natively solve the angular
momentum conserving form of the equations in a rotating frame, 
we implement all the terms on the right side 
of equation \eqref{eq:f2} as source terms. 
Though implementing coriolis forces in this manner
has been shown to produce inaccurate results for long-term integrations \citep{Kley1998}, for simplicity
we adopt the source 
term method and restrict our simulations to tens of orbits which is 
sufficient to study flows near 
the planet operating on fast dynamical times. This method remains 
untenable for studying processes such as gap-opening. 

We solve the equations in a spherical coordinate
system $(r_\ast,\theta_\ast,\phi_\ast)$ centered on the star 
and placing the orbital 
plane at $\theta_\ast=\pi/2$. The angular frequency of the rotating
frame is set equal to the constant Keplerian frequency of the orbit. 
For planet with semi-major axis $a$ and star with mass 
$M_\ast$, this aligns $\bm{\Omega_0}$ with the polar axis of the
$(r_\ast,\theta_\ast,\phi_\ast)$ system and sets
${\Omega_0}=GM_\ast/a^3$.
Under this construction, the gravitational potential $\Phi$ takes the 
form:
\begin{multline}\label{eq:pot}
    \Phi = -\frac{GM_\ast}{r_\ast} 
    - \frac{GM_p}{\sqrt{r_\ast^2 + r_p^2
    - 2r_\ast r_p\sin\theta_\ast\cos\left(\phi_\ast-\phi_p\right)
    +\epsilon^2}}\\
    + \frac{GM_p\sin\theta_\ast\cos\left(\phi_\ast-\phi_p\right)}{r_p^2}
\end{multline}
where $M_p$ is the planet mass and $(r_p,\phi_p)$ 
denote the radial and azimuthal coordinates of the planet. $\epsilon$ 
is a gravitational softening length, added 
to prevent the gravitational acceleration from becoming disruptively 
large near the planet. The third
term in equation \eqref{eq:pot}, known as the indirect potential, 
is included for the sake 
of completeness but results in unnoticeable differences for these 
models. 

To avoid the unnecessary computational complexity from
integrating Kepler's equations, we approximate the
eccentric orbital motion to first order in eccentricity $e$ and 
restrict the simulations performed to $e\leq 0.1$. Under 
this approximation the planet coordinates in the rotating 
frame take the form,
\begin{align}
    r_p &= a\left(1-e\sin\Omega_0t\right)\label{eq:r}\\
    \phi_p &= -2e\cos\Omega_0 t\label{eq:phi}.
\end{align}
With the aforementioned choice of $\bm{\Omega_0}$, this eccentric 
motion reduces to epicycles 
around the fixed point at $r_\ast=a$, $\phi_\ast = 0$ in the 
rotating frame.

\subsection{Length Scales and Model Parameters}
In the case of a planet orbiting a star and interacting 
hydrodynamically and gravitationally with a gaseous protoplanetary disk, 
there exist several established length scales characterizing the 
relevant physics. As we refer to these
quantities heavily in this work, it is prudent to define and discuss
them here. First, the planet's Bondi radius:
\begin{equation}
R_b \equiv \frac{GM_p}{c^2},
\end{equation}
defined here in terms of Newton's gravitational constant $G$, 
planet mass $M_p$ and sound speed $c$, arises as the natural length 
scale comparing the planet's gravity to the thermal state of the 
nebular gas. Similarly, the planet's Hill radius:
\begin{equation}
R_h \equiv a\left(\frac{q}{3}\right)^{1/3},    
\end{equation}
defined here in terms of the planet semi-major axis $a$ and the
planet-to-star mass ratio $q$, arises as the natural length scale 
comparing the strength of planet's gravity to that of its host star.
With $r_b$ as a length scale set by the planet-disk
interaction and $r_h$ as a scale set by the planet-star
interaction, it is natural to also consider some scale set by the 
star-disk interaction -- namely the disk scale height:
\begin{equation}
    H\equiv \frac{c}{\Omega},
\end{equation}
defined here as the ratio of sound speed to orbital frequency $\Omega$.
Because only two of these three scales are 
independent it is sensible to uniquely characterize a 
model by two parameters. For this work we 
use the planet's thermal mass $q_t$ defined by
\begin{equation}
q_t \equiv q\left(\frac{H}{a}\right)^3 = \frac{R_b}{H}
\end{equation}
as one of these dimensionless parameters. 
Choosing $q_t$ as the dimensionless parameter offers unique 
advantages over other potential options. 
In particular, it categorizes the dimensional length scales of the 
problem into the following well-defined hierarchy:
\begin{align}\label{eq:hier}
R_b < R_h < H &\quad\text{ for }\quad q_t < \sqrt{\frac{1}{3}} \\
R_h < R_b < H &\quad\text{ for }\quad \sqrt{\frac{1}{3}} < q_t < 1 \\
R_h < H < R_b &\quad\text{ for }\quad 1 < q_t < 3 \\
H < R_h < R_b &\quad\text{ for }\quad q_t > 3
\end{align}
For the ordering in equation \eqref{eq:hier}, where the thermal 
mass is small (i.e. subthermal), $q_t$ reduces to the only 
parameter necessary to 
characterize the problem \citep{Machida+2008}. 
We ensure that our models cover this subthermal regime by simulating 
$q_t=0.25$ and incrementing $q_t$ until computational 
challenges prevent us from obtaining interpretable 
results $(q_t\approx 2)$.
In order to accurately capture as wide a range of $q_t$ as possible,  
these simulations were designed to be global in nature. 
In the global framework, uniquely specifying a model requires a 
second dimensionless parameter which we choose to be the 
disk aspect ratio evaluated at the planet's semi-major axis:
\begin{equation}
    h_p = \frac{H}{a}
\end{equation}
Because these types of simulations are predominantly 
dependent on the choice of $q_t$ and not
$h_p$ \citep{Fung+2015} so we fix this secondary parameter 
to $h_p=0.05$ for all models.

The crux of this work lies in introducing a third dimensionless 
parameter -- the eccentricity $e$ -- and determining how this new 
parameter alters circumplanetary flows. The eccentricity is chosen 
to investigate a wide range of behaviors as the 
planet goes 
from circular ($e=0$) to subsonic eccentric motion ($e\lesssim 0.05$) 
to supersonic eccentric motion ($e\gtrsim 0.05$).

An additional parameter, the 
gravitational softening length $\epsilon$ though not physically 
motivated, is necessary to avoid computational difficulties. 
The softening length can have a dramatic effect on planet scale flows 
and can prevent CPDs from forming entirely by suppressing rotational 
velocities \citep{Fung+2019}. 
In a physical sense, one desires to make 
the softening length as small as the planet's physical radius 
while still large enough
to avoid any computational troubles. To this end, we run a set of 
simulations with small $\epsilon = 0.015 R_b$ which is small enough 
to form a flattened CPD around the planet. To verify the robustness of 
any results we run an additional set of simulations with large 
$\epsilon = 0.1R_b$ which is too large to form a flattened disk. 
We summarize the range of parameters explored in Table
\ref{tab:model} and provide labels to uniquely designate each model. 
The full suite of models covers all permutations of 
the parameters listed and future references to particular models
are indexed by their parameter labels. For example, $\texttt{q025e1}$ 
refers to model with $q_t = 0.25$, $h_p = 0.05$, $e=0.1$, and 
$\epsilon=0.015R_b$. All models are run for $20$ orbits as this tends 
to be enough time for flow fields to reach steady state 
(or quasi-steady state for eccentric models).

\begin{deluxetable*}{ccc}
\tablenum{1}
\tablecaption{Summary of Model Parameters \label{tab:model}}
\tablewidth{0pt}
\tablehead{\colhead{Parameter} & \colhead{Value} & \colhead{Label}}
\startdata
$q_t$ & $0.25$, $0.5$, $1.0$, $2.0$ & \texttt{q025}, \texttt{q05}
 \texttt{q1}, \texttt{q2}\\
$h_p$ & $0.05$ & -- \\
$e$ & $0$, $0.025$ $0.05$, $0.075$ $0.1$ & \texttt{e0}, \texttt{e025}, 
\texttt{e05}, \texttt{e075} \texttt{e1}\\
$\epsilon/R_b$ & $0.015$, $0.1$ & -- , \texttt{hs}\\
\enddata
%\tablecomments{}
\end{deluxetable*}

\subsection{Computational Details}
\subsubsection{Initial Conditions}
We initialize the disk in axisymmetric hydrostatic equilibrium at time $t_0$
with the same density and velocity profiles as \citet{Fung+2019}. 
In the rotating frame they are:
\begin{equation}\label{eq:rho0}
    \rho\left(t_0\right) = \rho_0 \left(\frac{r_\ast\sin\theta_\ast}{a}\right)^{-3}
    \exp{\left[-\frac{GM_\ast}{r_\ast c^2}
    \left(\frac{1}{\sin\theta_\ast}-1\right)\right]}
\end{equation}
\begin{equation}\label{eq:v0}
    \bm{v} = \left(\sqrt{\frac{GM_\ast}{r_\ast\sin\theta_\ast}
     -3c^2} 
    -r_\ast\Omega_0\sin\theta_\ast\right)\bm{\hat{\phi}_\ast}
\end{equation}
We set $\rho_0=1$ because the self-gravity of the gas is ignored. 
This initial 
profile corresponds to a vertically integrated gas surface density that 
scales as $r_\ast^{-3/2}$. We also introduce the component of the 
potential due to the planet $\Phi_p$ gradually with the following form:
\begin{equation}
   \Phi_p\left(\bm{r},t\right) 
   =\Phi_p\left(\bm{r}\right)\sin^2\left(\frac{\Omega_0 t}{8}\right).
\end{equation}
This form increases the potential from $0$ at $t=0$ to its full value 
after $2$ orbits.
\subsubsection{Domain and Boundaries}
For all models the same domain is used. The grid extends from 
$3a/5$ to $5a/3$ in radius with a logarithmic spacing. 
For radial boundary conditions, the fluid variables in the 
ghost cells are fixed to the equilibrium values obtained from equations
\eqref{eq:rho0}-\eqref{eq:v0}. Our refinement prescription for the mesh 
(see \ref{sec:Res}) introduces sufficient diffusion near the 
boundaries to render wave-killing boundaries unnecessary.
Because the equations have vertical symmetry across the 
midplane, we save a factor of $2$ in computation by 
modeling only half the domain in $\theta_p$. 
The polar grid is then set only from $\theta_p=\pi/2-4H_p$ 
to $\pi/2$ with a linear mesh spacing and reflecting boundary 
conditions on both boundaries. The azimuthal grid covers the full 
$2\pi$ with linear mesh spacing. With this domain, the root grid 
is set to have $64$ cells in $r_\ast$, $16$ cells in $\theta_\ast$, 
and $512$ cells in $\phi_\ast$ to keep cells roughly square.

\subsubsection{Resolution}\label{sec:Res}
To study flows near the planet within a global simulation of 
the full protoplanetary disk necessarily requires 
a refinement scheme to concentrate resolution on the scales of 
interest. \textsc{Athena++} comes equipped with
a block-based adaptive refinement scheme that can be leveraged for 
these types of simulations. In the block-based approach, meshblocks 
made up of grid cells are refined in a self-similar manner based on 
some user-defined criteria. For our criteria, we set a minimum cell 
size desired over a control volume centered on the planet and 
allow the mesh to adjust itself to meet these requirements.
Though this means the resolution cannot be controlled in regions 
outside the specified control volume, 
the number of cells per meshblock can be tuned to either relax or 
steepen how quickly the resolution changes as one moves away from the 
planet. For our fiducial resolution, the minimum cell width is set
to $W_{\min} = R_b/256$ within a control volume of $N_\text{vol} = 32$ 
cells in each direction from the planet.
This corresponds to a resolution of $1/W_{\min} = 256$ cells per $R_b$ 
in the innermost $r < R_b/8$. The meshblock size is fixed at 
$N_{i} = 16$ cells for each direction. 
With this refinement prescription, the cell width $W_i$ in the
the $i$ direction at coordinate 
$x_i\in\left[r_\ast,\theta_\ast,\phi_\ast\right]$ 
is approximately,
\begin{equation}
W_i= \begin{cases} 
      W_{\min} & \left|x_i-x_p\right| < N_\text{vol}W_{\min}  \\
      2^b W_{\min}  & \left|x_i-x_p\right| > N_\text{vol}W_{\min}
 \end{cases}
\end{equation}
\begin{equation}
b \equiv \left\lfloor \log_2
\left(1+\frac{\left|x_i - x_p\right|-N_\text{vol}
W_{\min}}{2N_iW_{\min}}\right) \right\rfloor
\end{equation}
where $x_p$ is one of $\left[r_p,\theta_p,\phi_p\right]$ depending on 
the choice of $i$. To verify the accuracy of the fiducial models,
higher resolution simulations are also performed. The high 
resolution models simply double the resolution everywhere in the 
domain. The fiducial models with $\epsilon=0.015R_b$ are not fully 
converged near the planet for some quantities like the integrated mass
(Section \ref{sec:den}). 
In these cases, the converged hydrostatic models with large $\epsilon$ 
are used to inform our understanding. 
Nevertheless, we find that the character of 
flow and conclusions presented are robust with respect to all models 
as the resolution is increased.

\subsection{Frames and Coordinates}
In Section \ref{sec:eqns}, 
we introduced a spherical coordinate system for the 
non-inertial rotating frame centered on the star in which we perform 
our simulations. In this computational frame, $e=0$ planets lie fixed 
at the point 
$\left(r_\ast,\theta_\ast,\phi_\ast\right) = \left(a,\pi/2,0\right)$ 
while $e>0$ planets follow a periodic epicyclic path about the
same point. 
Because we are predominantly interested in the character of flows
near the planet, in the following analysis it becomes advantageous to 
work in a coordinate system centered on the planet 
rather than the star. We construct a Cartesian system on the planet 
indexed by coordinates $(x,y,z)$. We align this new frame to be 
curvilinear with respect to the inertial motion of the planet. In 
particular, in an inertial frame, $\bm{\hat{y}}$ is chosen to always be 
aligned with the instantaneous direction of orbital motion. 
$\bm{\hat{z}}$ is chosen to lie perpendicular to the orbital plane 
(by this design it points in the same direction as the 
star-centered polar axis).  Finally, $\bm{\hat{x}}$ is chosen to ensure 
the usual right-handedness of Cartesian $\left(x,y,z\right)$. We 
will also frequently refer to this frame using spherical coordinates
to discuss the rotational and radial aspects of the flow.
The spherical coordinates are defined with the usual 
Cartesian-spherical transformation:
\begin{align}
x&=r\sin\theta\cos\phi \label{eq:sph0}\\
y&=r\cos\theta\cos\phi\\
z&=r\cos\theta \label{eq:sph2}
\end{align}
and are unscripted $\left(r,\theta,\phi\right)$ to distinguish 
them from the star-centered 
coordinates $\left(r_\ast,\theta_\ast,\phi_\ast\right)$. With the same 
coordinate transformation, equations \eqref{eq:sph0}-\eqref{eq:sph2},
we also define a set of Cartesian coordinates 
for the frame centered on the star $\left(x_\ast,y_\ast,z_\ast\right)$

\section{The Circular Orbit}\label{sec:e0}
There already exist a number of studies investigating the interaction 
of the circular orbit planet with its isothermal environment in 3D 
\citep{Machida+2008,Ormel+2015,Fung+2015,Kurokawa+2018,
Kuwahara+2019,BethuneRafikov2019}.
Within this body of literature and across various 
numerical implementations 
(e.g. \textsc{PEnGUIn}\citep{Fung+2014} vs. 
\textsc{PLUTO}\citep{Mignone+2007} vs. 
\textsc{Athena++},
spherical vs. Cartesian coordinates, global vs. local approximation), 
several key aspects of the 
flow remain consistently observed. 
We demonstrate here that, in the limit of a circular orbit, our results are consistent with theirs.

\subsection{Density}
\begin{figure*}
\gridline{\fig{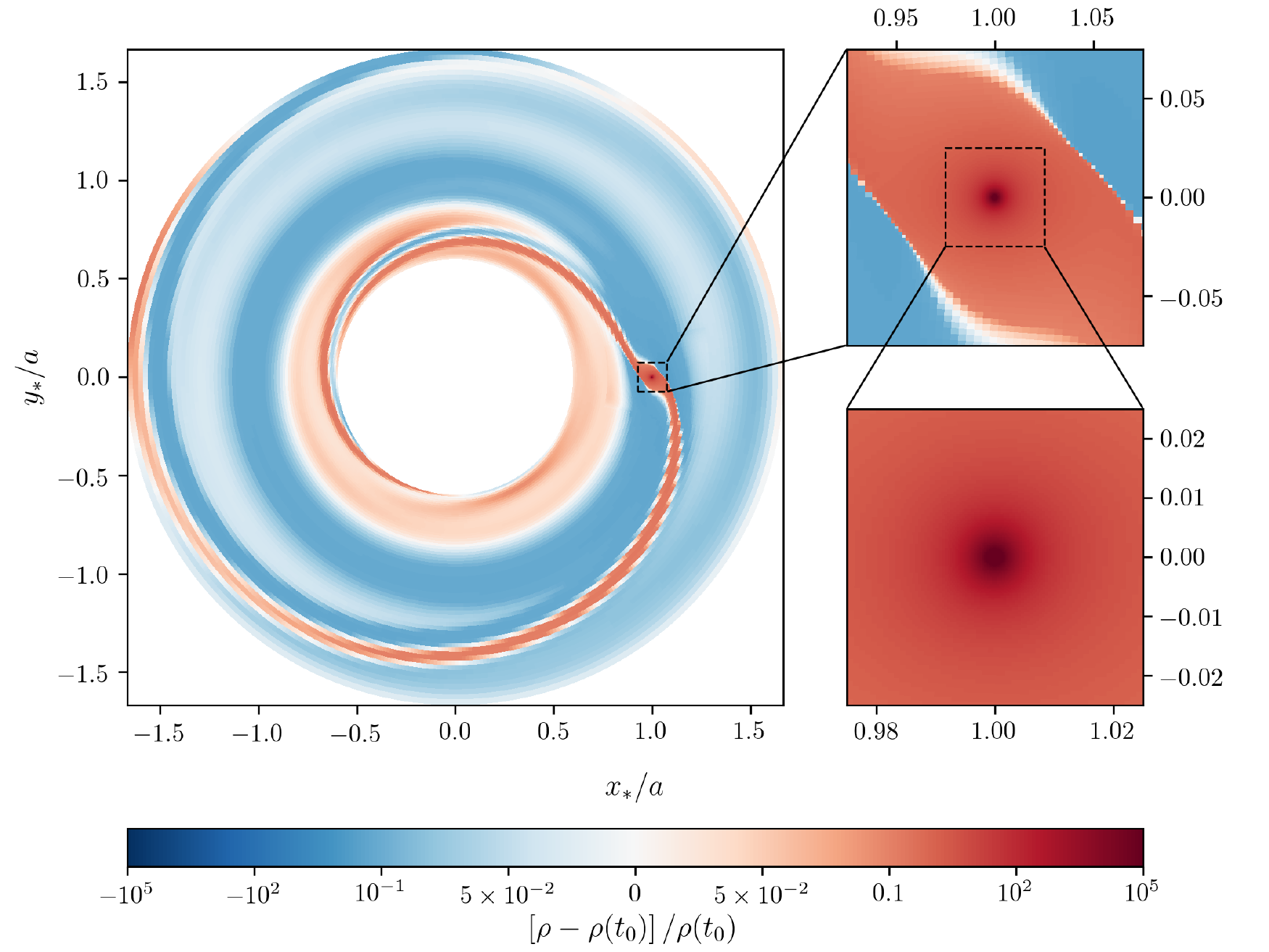}{\textwidth}{}}
\caption{
The steady state density perturbation in the orbital plane for the
circular orbit model with $q_t=1$. The panels progressively 
zoom from the full disk into a $1.5R_b\times 1.5R_b$ region and 
$R_b/2\times R_b/2$ region centered on the planet.
The colormap is chosen to highlight small amplitude global 
perturbations and the dramatic exponential density profile near 
the planet.}
\label{fig:1}
\end{figure*}

\begin{figure*}
\gridline{\fig{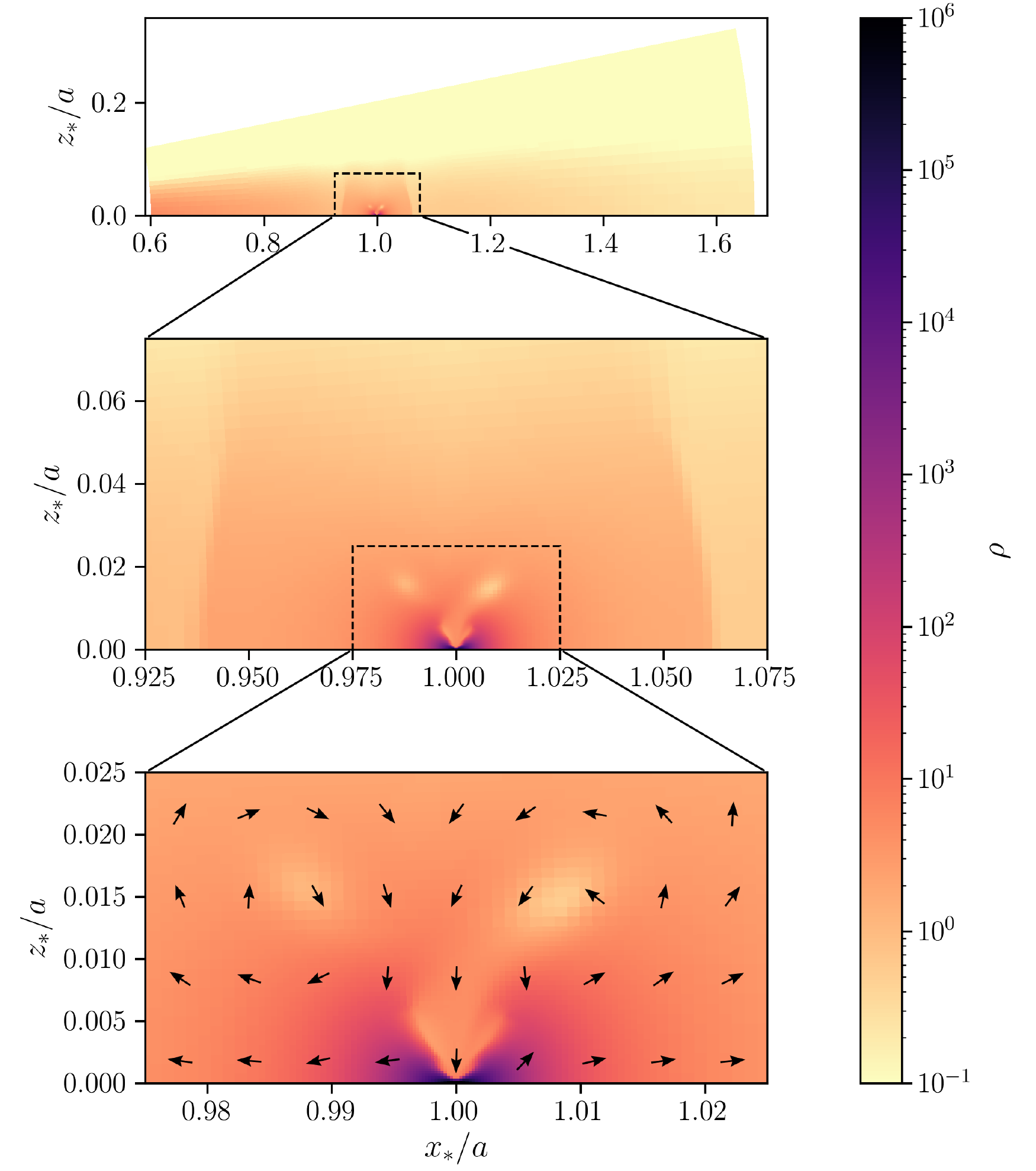}{0.98\textwidth}{}}
\caption{
A $\phi_\ast = 0 $ slice of the steady state density profile for the
circular orbit model with $q_t=1$. The panels progressively 
zoom from the full disk into a $1.5R_b\times 1.5R_b$ region and 
$R_b/2\times R_b/2$ region centered on the planet. Fixed length 
arrows point in the direction of the poloidal velocity
$\bm{v\cdot\hat{r}_\ast}+\bm{v\cdot\hat{\theta}_\ast}$ and 
highlight the meridional circulation.}
\label{fig:2}
\end{figure*}

Figure \ref{fig:1} highlights the midplane density perturbation 
caused by the introduction of a planet, using model \texttt{q1e0} as 
an example.
On the global scale, we find the expected spiral waves 
\citep{OgilvieLubow2002} and a shallow gap forming at the 
planet's orbit.
As one zooms into the 
region near the planet, the density begins to increase rapidly, owing 
to the isothermal equation of state.
Whereas the \texttt{hs} models take 
on the character of a hydrostatic symmetric sphere in the 
region $r\lesssim R_b/2$, 
the small $\epsilon$ models have non-negligible rotational support 
and a vertically flattened density structure near the planet. 
The steady state vertical density profile for \texttt{q1e0} 
is displayed in Figure \ref{fig:2}. Zooming into the inner $R_b/2$ 
near the planet in the third panel, one can see the existence of 
low density polar region and
a flattened high density CPD near the planet. This agrees with the 
work of \citet{Fung+2019} where isothermal CPD are shown to be 
ubiquitous given a small enough $\epsilon$. 

A feature of these models, that we have not found 
sufficient reference to in most existing works 
is the presence of two low density vortices in the polar region
above the planet. There is some evidence for these vortices 
in \citet{Fung+2015} and \citet{Fung+2017}.
They find vortices occurring at approximately the same 
location but weaker depending on the choice of code. 
Here they are more dramatic, perhaps owing to the smaller choice 
of $\epsilon$. Because vortices create pressure minima at their core, the isothermal EOS produces corresponding density minima. As a result, 
these vortices can be seen in 
Figure \ref{fig:2} as the sliced low density lobes 
appearing near $\left(x_\ast,z_\ast\right) = \left(0.99a,0.015a\right)$ 
and $\left(x_\ast,z_\ast\right) = \left(1.01a,0.015a\right)$. 
Though these vortices exist for all models with small $\epsilon$, 
they become more tenuous as $q_t$ decreases.

\subsection{Flow Field}
\begin{figure*}
\gridline{
          \fig{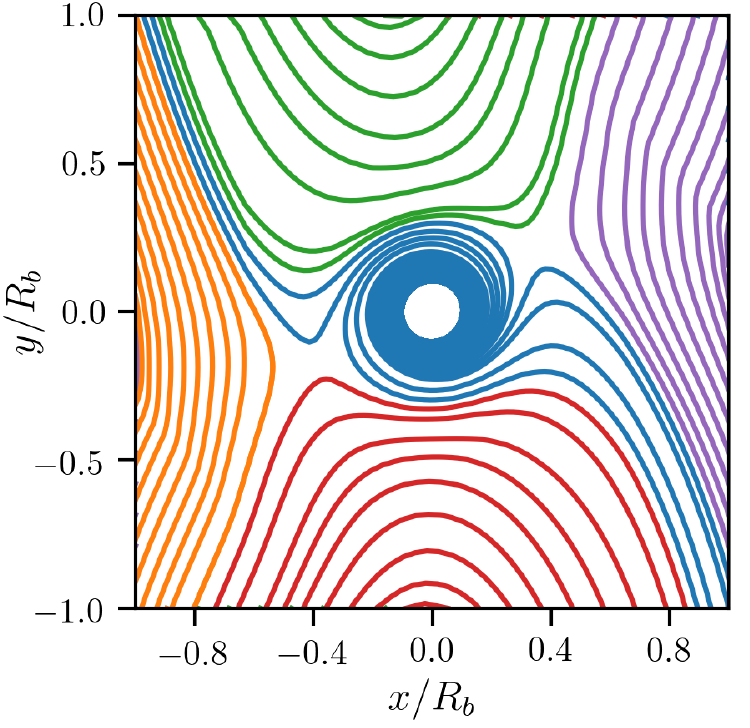}{0.5\textwidth}{}
          }
\caption{Steady state midplane streamlines in the circular case -- 
here from model \texttt{q1e0}. Green and red lines mark horseshoe 
streamlines. Purple and orange lines mark the background disk streamlines. 
Blue marks the outflowing recycling streamlines sourced from higher 
latitudes.
}
\label{fig:stream}
\end{figure*}

The most salient point to emerge from circular orbit models is
a pattern of meridional circulation near the planet. Both the models 
here and in other works
\citep{Machida+2008,AyliffeBate2012,Tanigawa+2012,DangeloBodenheimer2013,
Ormel+2015,Fung+2015,Szulagyi+2016,Fung+2017,Szulagyi2017,
LambrechtsLega2017,Cimerman+2017,Kurokawa+2018,Kuwahara+2019,
BethuneRafikov2019,Schulik+2019}
find that the flow near the planet is one in which gas flows in 
through the poles and out through the midplane. This finding is, in 
turn, used to inform full evolutionary models of planet and satellite
formation \citep{BatyginMorbidelli2020}. An example of the 
meridional circulation in these models is shown in 
Figure \ref{fig:2} as arrows in the lowermost panel. The same pattern 
of meridional circulation is present for all $q_t$. 
The circulation is fed from high latitude horseshoe orbits which then 
descend toward the planet. 
The polar inflow then 
reaches supersonic speeds, once the gas descends to within 
$\sim R_b/2$ of the planet. Near the midplane the flow turns 
and spirals outwards exiting near the planet's Lagrange points. This 
can be seen in Figure \ref{fig:stream} where we plot representative 
midplane streamlines for ciruclar orbit model \texttt{q1e0}. 
The blue curves correspond to the aforementioned outflows 
sourced from high latitude horseshoe orbits.
The other families of midplane flows are also pictured in Figure 
\ref{fig:stream} with 
horseshoe flows in red and green and the background disk flows in 
purple and orange. Together, these three families of streamlines 
define the well established pattern of midplane flow for 
3D isothermal simulations.

The same meridional circulation prevents solids in the midplane from 
being readily accreted due to the outflowing nature of the gas there.
Because solids tend to settle to the midplane, 
this so-called outflow barrier \citep{Kuwahara+2019}, could
insulate the planet from a significant reservoir of solids. 
This behavior has been confirmed in 3D simulations of pebble accretion 
\citep{KuwaharaKurokawa2020} 
where midplane particles with dimensionless Stokes number
$\tau_s\lesssim q_t^2$ are partially prevented from accreting.

The flow field of our circular models is more explicitly provided
in the next section where we make direct comparison with the eccentric 
models.

\section{Comparison with the Eccentric Orbit}\label{sec:ecc}
Within several orbits, eccentric models reach a quasi-steady state 
where the state is approximately periodic with period $2\pi/\Omega_0$.
Because the states of the eccentric models are the same modulo 
one orbit, it is simplest to refer to a particular time in a 
quasi-steady model by the phase of its orbit $\psi$. 
In what follows, our convention uses $\psi=\pi/2$ as the phase of 
periapse and $\psi=3\pi/2$ as the phase of apoapse. 

\subsection{Density}\label{sec:den}
While the eccentric motion is still subsonic
$\left(e\lesssim 0.05\right)$, the density field in the eccentric 
models is not appreciably altered from the circular profile.
When the eccentric motion becomes supersonic 
$\left(e\gtrsim 0.05\right)$, a bow shock appears, making the density 
field a function of orbital phase.
In the subthermal models, this bow shock tends to increase the overall mass and density by increasing the pressure downstream of the shock.

\begin{figure*}
\gridline{
          \fig{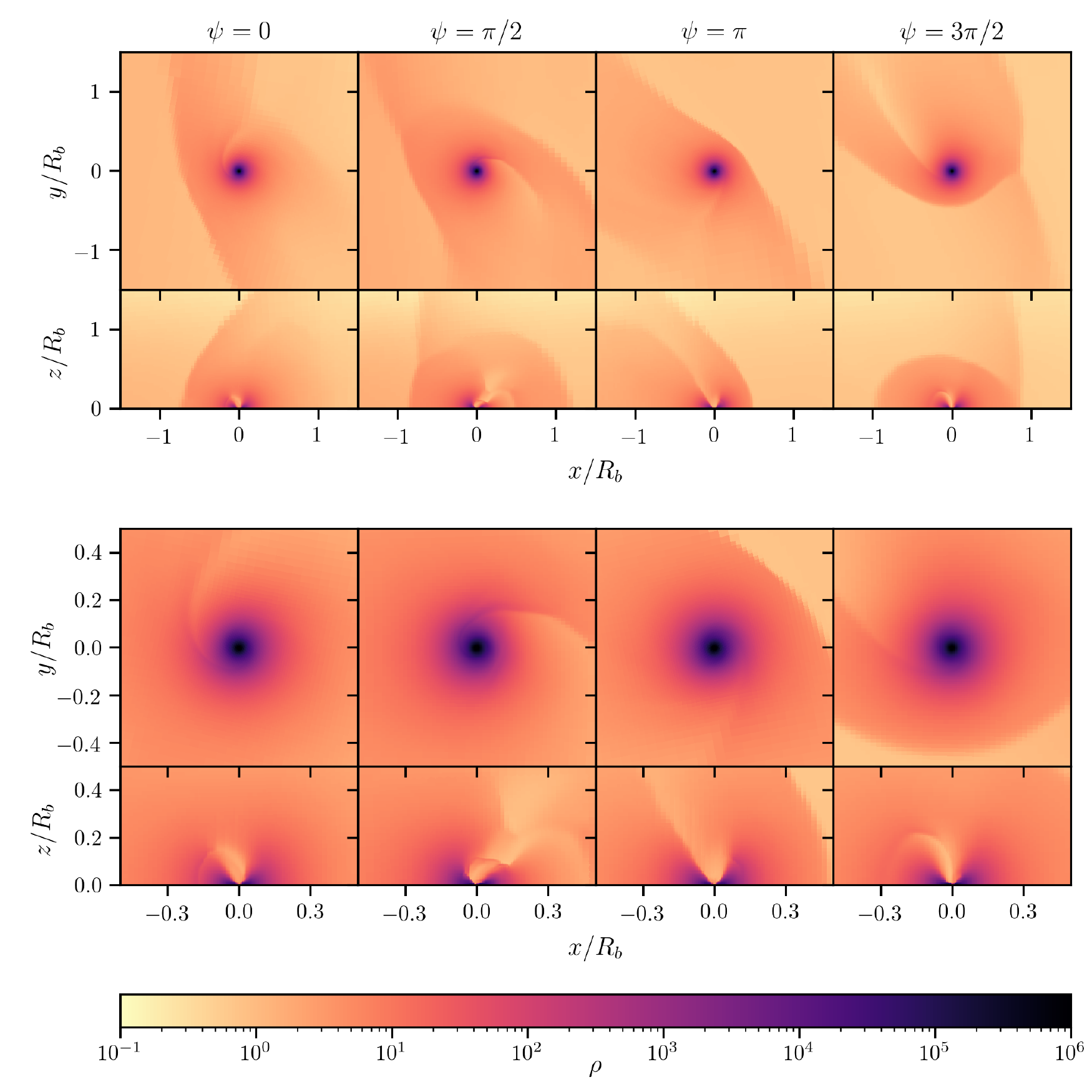}{\textwidth}{}
          }
\caption{The midplane $\left(\theta_\ast=\pi/2\right)$ and 
vertically sliced $\left(\phi_\ast=0\right)$ density field for model 
with $q_t=1$ and $e=0.1$ at various phases of orbit. The top panel 
encompasses the full Bondi radius ($r<1.5 R_b$) while the 
bottom is zoomed to show the inner $r<0.5R_b$.}
\label{fig:dene}
\end{figure*}

This phase-dependent density field is apparent in the midplane 
and vertically sliced density profiles provided in 
Figure \ref{fig:dene}. 
The bow shock can be seen in the $q_t=1$ models particularly at 
$\psi=3\pi/2$ (Figure \ref{fig:dene}) but because the flow is 
already sonic near $R_b$ in those circular models, the bow shock is 
not nearly as robust or influential as in lower $q_t$ models.
The bow shock forms between $R_b/2$ and $R_b$ and is nearly
perpendicular to the direction of eccentric motion, meaning its 
orientation rotates a full $2\pi$ around the planet over the course 
of one orbit. 

\begin{figure*}
\gridline{
          \fig{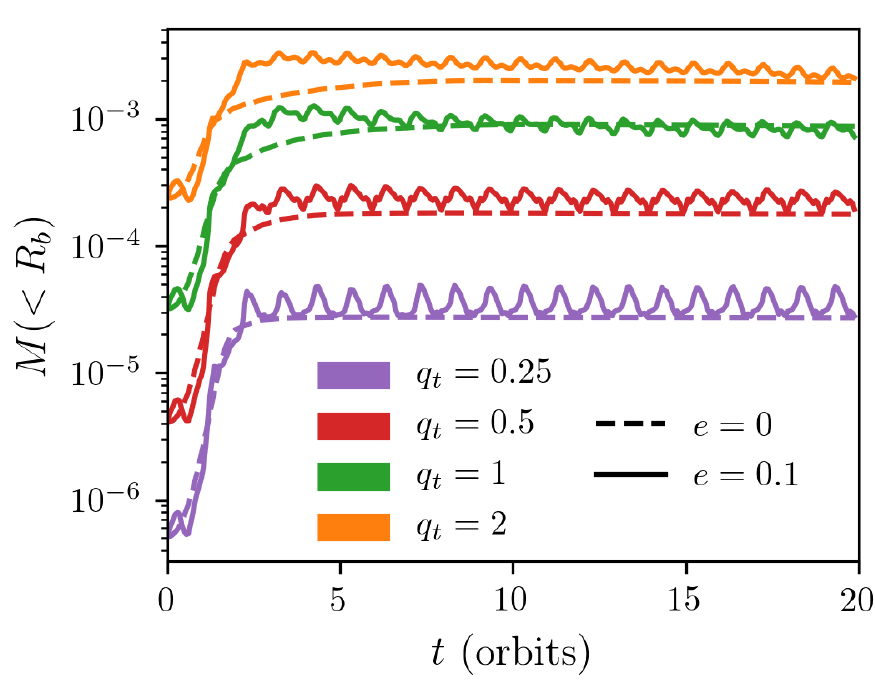}{0.5\textwidth}{}
          }
\caption{Total mass interior to $R_b$ as a 
function of time for the \texttt{hs} models.}
\label{fig:bulk}
\end{figure*}

In Figure \ref{fig:bulk}, we plot the total mass interior 
to $R_b$ as a function of time for the \texttt{hs} models.
The masses for the \texttt{e1hs}
planets tend to track the circular cases with some additional 
periodic variation, but in the $q_t=0.25$ and $0.5$ cases, the masses 
of these low $q_t$ planet atmospheres show an average uniform systematic 
enhancement of $\sim 40\%$. 
We use the \texttt{hs} models here in particular because the mass 
measurements are not entirely converged for smaller $\epsilon$. 
The mass in the unconverged cases is dominated by a contribution close to the planet ($r<\epsilon$), which we cannot properly resolve even with several additional levels of mesh refinement.

\subsection{Flow Field}

\subsubsection{Rotation}\label{sec:rot}
In the circular case, gas well inside $R_b$ rotates in a prograde sense, and gas on the scale of $R_b$ merges with the background shear (which has a negative, or retrograde, sense of ``rotation'', even though it does not circulate the planet.)
In the eccentric case, we find that the rotation of the CPD ($r\lesssim 0.05 R_b$) is not significantly affected by eccentricity. 
In both the circular and the eccentric cases, the CPD appears to 
remain bound. Though determining what gas is 'bound' in these types 
of simulations is difficult, the CPD gas is bound in a kinematic 
sense. By this we mean: orbit-averaged midplane radial 
velocities are directed towards the planet within the CPD and 
away from the planet exterior to the CPD. This result is remarkably consistent with those of \cite{Fung+2019}.

Exterior to the CPD, the rotation rate appears sensitive to eccentricity when the planet is subthermal. In this parameter space, eccentricity, while still subsonic, may increase prograde rotation, but it rapidly flips the rotation to retrograde once it crosses into supersonic.

\begin{figure*}
\gridline{
          \fig{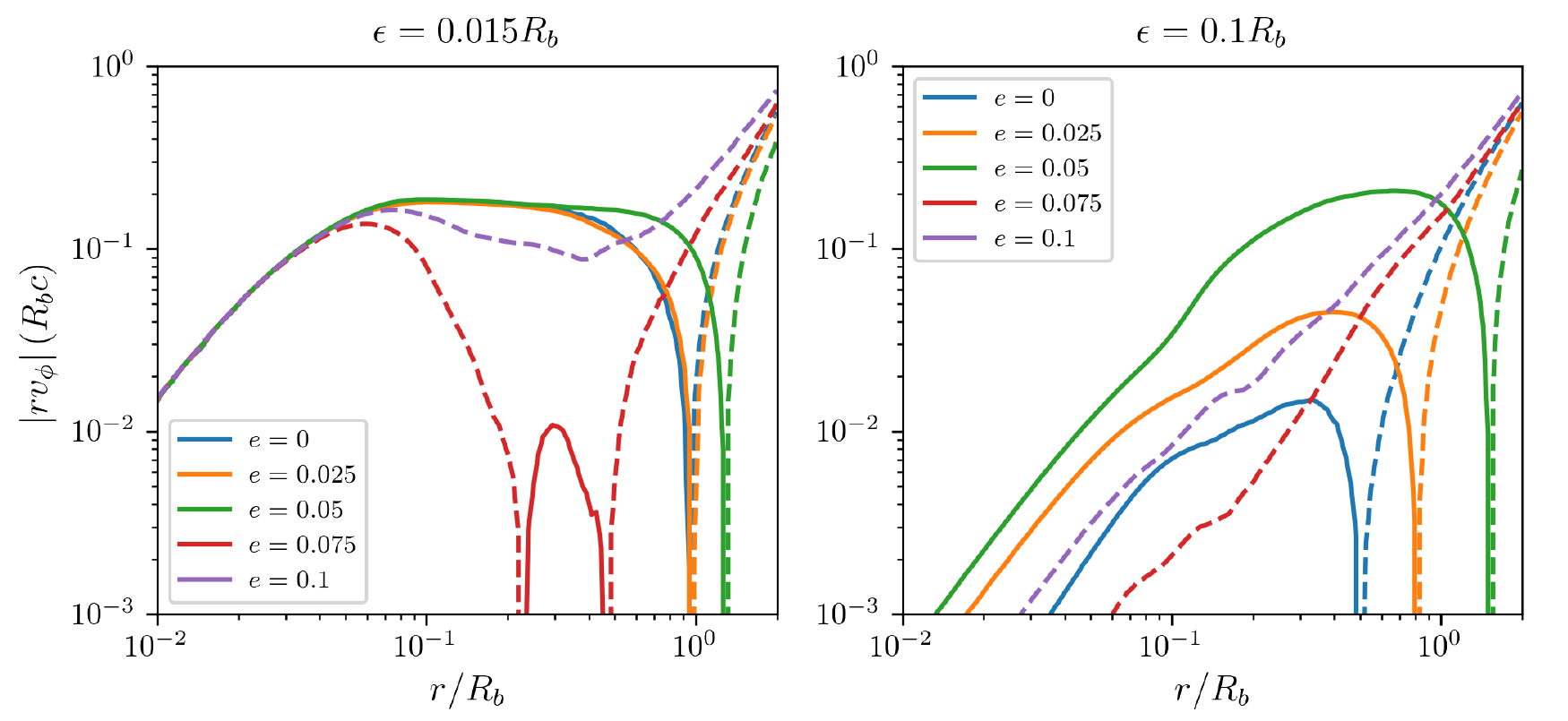}{\textwidth}{}
          }
\caption{Magnitude of the orbit averaged midplane specific 
angular momentum for models with $q_t=0.25$. The left panel shows the 
fiducial model rotation curves whereas the right panel shows the \texttt{hs} 
models. Solid lines plot positive (prograde) rotation whereas dashed lines 
plot negative (retrograde) rotation.
}
\label{fig:rot}
\end{figure*}

We say subsonic eccentric motion \textit{may} increase prograde 
rotation because we only detect significant spin-up in one of our models, \texttt{q025e05}, which also has the lowest planet mass in our parameter space.
To understand this behavior, consider the 
acceleration the gas experiences due solely to the eccentric motion. 
In the frame of the planet, and to first order in $e$, this acceleration
takes the form
\begin{equation}\label{eq:force}
\bm{a_e} = 2a\Omega_0^2 e\left(\sin\Omega_0 t\bm{\hat{x}}
-\cos\Omega_0t \bm{\hat{y}}\right) 
\end{equation}
of forced harmonic motion. Gas in the circular case that orbits the planet 
at angular velocity $\Omega$ greater than $\Omega_0$, experiences the 
orbit averaged acceleration. Because the gas on the scale of the CPD does
rotate with $\Omega \gg \Omega_0$ and the orbit averaged $\bm{a_e}$
vanishes, the rotation curve of the CPD remains unchanged between 
circular and eccentric models. Further out from the planet, $\Omega$ 
drops and the harmonic forcing of Equation \eqref{eq:force} can 
operate on the gas in nontrivial ways. 
This case of 
spin-up can be seen in Figure \ref{fig:rot} where we present the 
orbit averaged specific angular momentum as a function of radius for 
the \texttt{q025} models.

Though the spin-up occurs for only one fiducial model, we suspect 
this behavior is more generic at lower $q_t$ and that its limited 
observance here is due only to our choice of parameter space. 
This hypothesis is primarily motivated by the \texttt{hs} models 
where a pattern of spin-up is more prevalent (Figure \ref{fig:rot}). 
Although these models with a large softening length are more akin 
to planets with large radii relative to $R_b$ 
(i.e. super-Earths), they can nonetheless be used to inform 
trends about other regions of parameter space.
Of the \texttt{hs} models, \texttt{q025e025}, \texttt{q025e05}, 
and \texttt{q05e05} exhibit increased prograde rotation with the 
increase being greater as one goes to higher $e$ or lower $q_t$. 
This suggests the existence of some criterion as a function of $q_t,e$ 
that determines whether an eccentric model reproduces the circular 
rotation curve or whether it shows an enhanced level of rotation.
We suspect that a similar criterion 
exists for the fiducial, smaller $\epsilon$ models, 
but is more stringent due to the 
increased gravity. If the fiducial models are following the same 
pattern as the \texttt{hs} models, then \texttt{q025e05} just happens 
to be the only spin-up model that falls into our searched parameter 
space and any models with $q_t<0.25$, $e=0.05$ should also have 
spin-up.
Testing of the fiducial circular models with a smaller value of 
$\epsilon$ returns the same rotation curve on the scale of $R_b$. 
Therefore, we conclude that the flow is converged on the scale where 
spin-up is occurring. Explicit confirmation 
of the spin-up ubiquity in lower $q_t<0.25$ models however remains 
inaccessible with our current numerical setup. 

In the supersonic case, the rotational behavior of the gas changes 
and the above discussion of spin-up does not apply.
For superthermal planets, the orbit averaged rotational velocities 
show slight enhancement on the scale of $R_b$ but interior to 
$0.1 R_b$, the circular rotation curves are reproduced, suggesting 
again that the CPD is unaltered.
For the subthermal models, gas within $R_b$ has a lower 
magnitude of rotation than the circular orbits with the \texttt{q025} 
models even rotating in the retrograde sense. 
This state of suppressed rotation is possibly due to gas losing 
angular momentum as it passes through the bow shock. It is uncertain 
whether the retrograde behavior 
continues for lower $q_t$ but the absolute value of 
the rotation curve in all models
tracks the circular CPD rotation precisely.

\subsubsection{Meridional Circulation}\label{sec:circ}

As the eccentricity increases, models get further from the 
established paradigm of meridional circulation. No longer is the 
planet's atmosphere a steady polar inflow that turns and exits 
through the midplane as in Figure \ref{fig:2}. The changes to the
flow in the eccentric case can be summarized as follows:
\begin{itemize}
    \item The flow is highly variable in both time and space.
    \item The eccentric flow is periodic with period $2\pi/\Omega_0$.
    \item The inflow and outflow can be larger in magnitude than 
    the equivalent circular orbit flows.
    \item Though there is never truly significant outflow through the 
    pole, the dominant source of inflow can actually be through the 
    midplane (instead of the pole for circular orbits).
    \item The earlier points only become more apparent as one goes 
    to model with lower $q_t$ or higher $e$. 
\end{itemize}
\begin{figure*}
\gridline{
          \fig{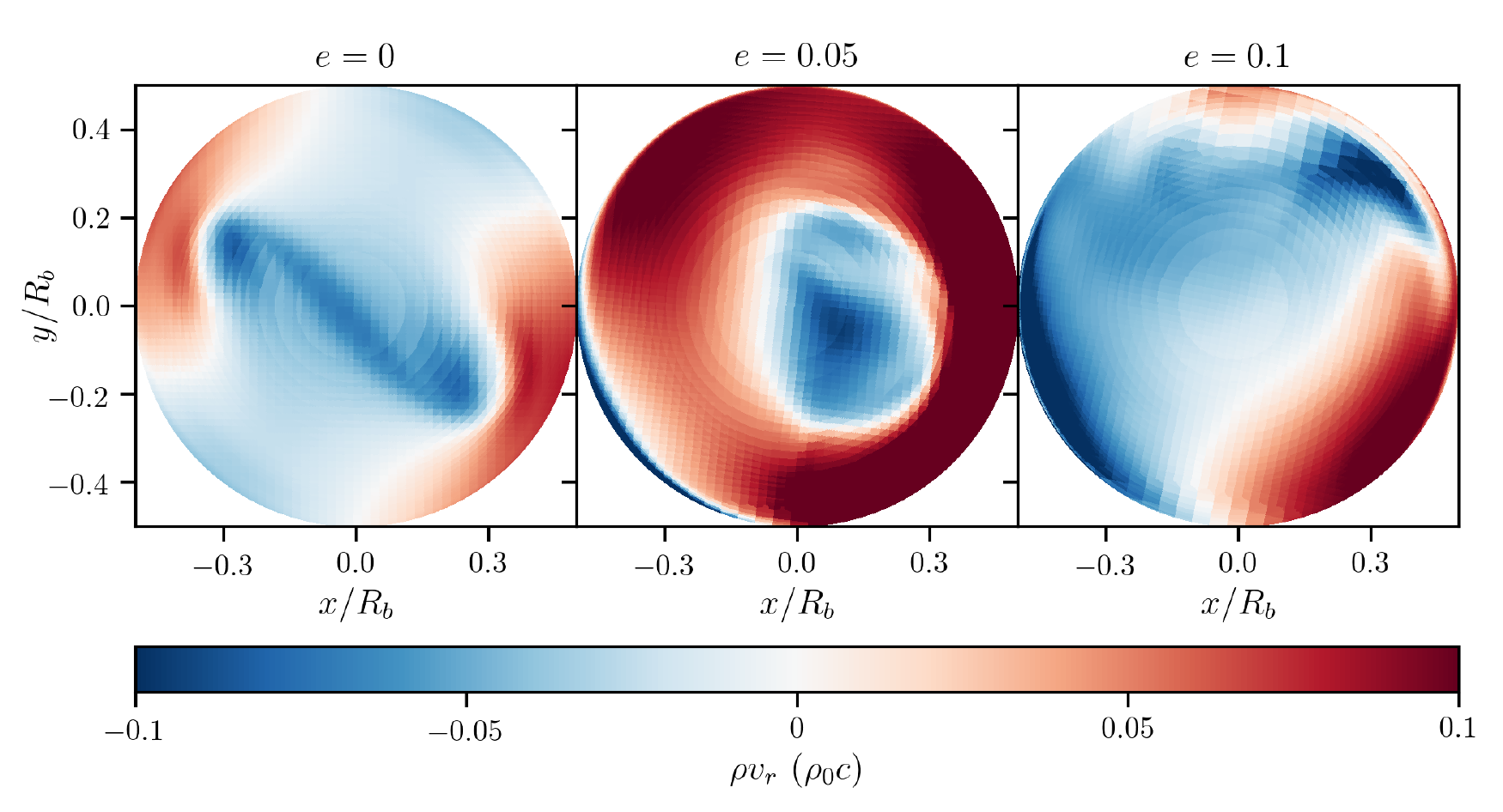}{\textwidth}{}
          }
\caption{The radial mass flux $\rho v_r$ through spherical surface with 
radius $r=R_b/2$ as a function of eccentricity. Thermal mass is $q_t=1$
and the planet is at $\psi =0$ in its orbit. Mass flux is projected 
onto the midplane hence the figure shows a top-down view.}
\label{fig:flux}
\end{figure*}
Figure \ref{fig:flux} provides a concrete example of the kind of 
variability exhibited between different models. The $e=0$ radial mass 
flux $\rho v_r$ exhibits the standard pattern of inflow/outflow with 
outflow occurring in the midplane primarily along $\phi=3\pi/4$ 
and $\phi=7\pi/4$. Adding a small $e$ simply moves the projected
mass flux pattern laterally in epicycles. 
As $e$ increases, the flow becomes highly variable. Orbit-averaging would lead to large cancellations between in- and out-fluxes.
In this sense, the orbit averaged mass flux is misleading---the 
amount of gas processed through the planet's Bondi radius can be 
considerably larger than that implied by the average mass flux.

To truly describe the flow in these eccentric models it becomes 
necessary to consider the inflow and outflow independently instead 
of just the average. 
\begin{figure*}
\gridline{
          \fig{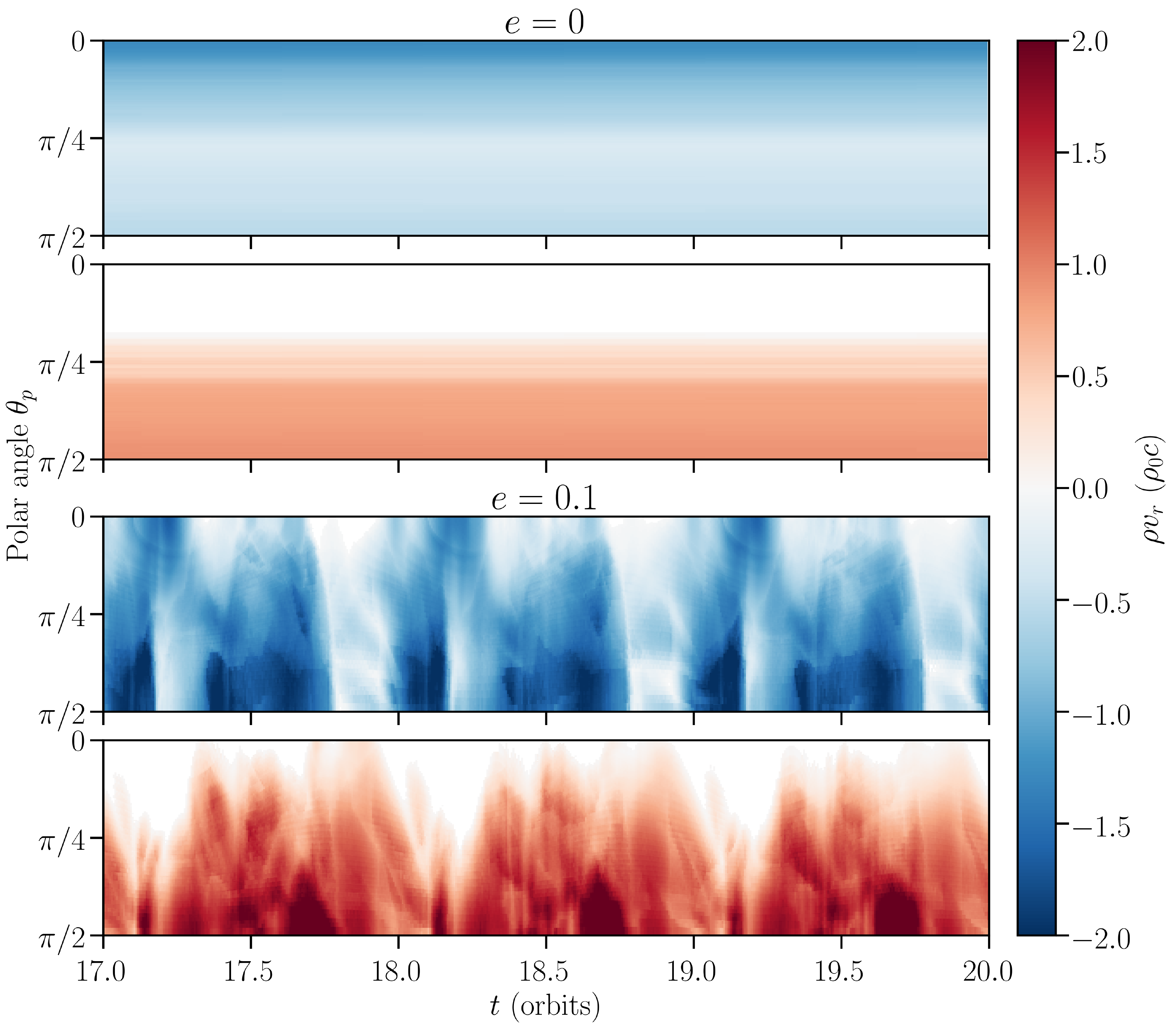}{\textwidth}{}
          }
\caption{Azimuthally averaged mass flux through sphere at $r=R_b/2$ 
model with $q_t=1$. The mass flux averages are organized according to
inflow (blue) or outflow (red).
Data is shown for 3 orbits after the $e=0$ model 
reaches steady state (top) and after the $e=0.1$ model reaches a 
periodic steady state (bottom).}
\label{fig:geom}
\end{figure*}
In Figure \ref{fig:geom}, this is done explicitly 
by averaging the inflowing and outflowing mass fluxes separately over 
azimuth. The result is a time series of the mass flux as a function 
of polar angle for a particular radius ($R_b/2)$. 
This Figure in particular highlights many of the
essential differences found between eccentric and circular orbits.
We see that the magnitude of the flow becomes highly variable and exhibiting periods of significant increase. The geometry is also 
changed, with the dominant inflow contribution actually coming 
from the midplane.

\begin{figure*}
\gridline{
          \fig{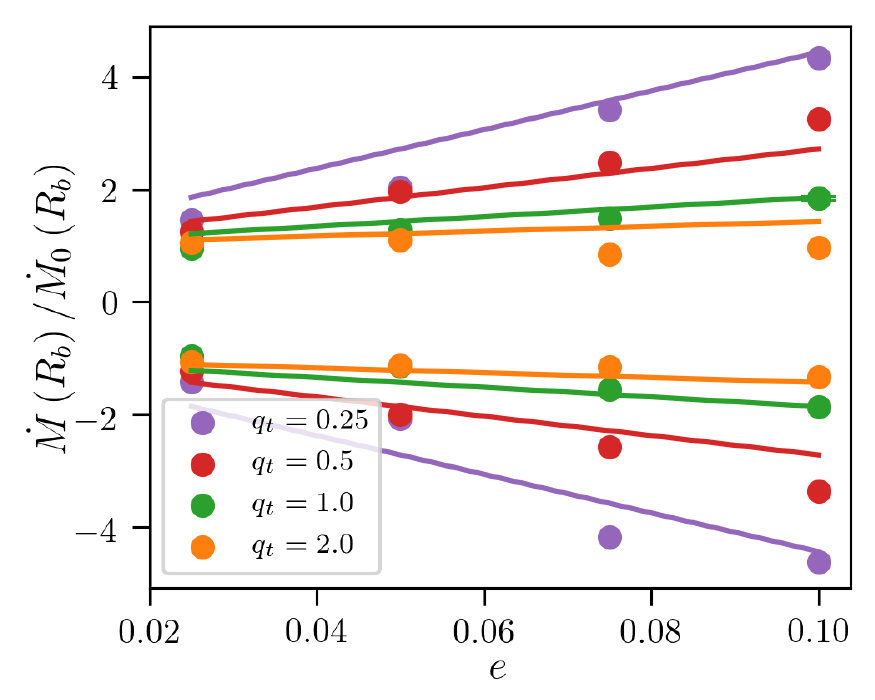}{0.5\textwidth}{}
          }
\caption{Mass inflow rate $\dot{M}$ as a function of $q_t$ and
$e$, normalized by the magnitude of $\dot{M}$ in the $e=0$ case. 
$\dot{M}$ is evaluated at $r=R_b$.}
\label{fig:mdot}
\end{figure*}

For models with higher $q_t$ or $e$, the inflow/outflow 
magnitudes increase, becoming several times larger than the circular 
model in some cases. This is shown in Figure \ref{fig:mdot} where we 
plot the amount of inflowing and outflowing mass through the Bondi 
radius averaged independently over the course of an orbit. 
These mass flow rates $\dot{M}$ are also normalized by their 
circular equivalents $\dot{M}_0$
to give the relative enhancement to the mass of processed gas as a 
function of our dimensionless model parameters $q_t$ and $e$.
We refer to this eccentric flux relative to the circular flux as 
the enhancement factor and find that it scales inversely proportional with $q_t$ and linearly with $e$. This scaling can be interpreted as follows.
 
If the velocity in eccentric models is set 
by the eccentric motion, it scales as $v\sim a\Omega_0 e$. 
Meanwhile, the velocities in the circular case at $R_b$ 
are set by the 
velocity of the background shear which is on the order of
of $v\sim \Omega_0 R_b\sim a\Omega_0 q_t h_p$ at $r=R_b$. 
Because the background densities are similar between eccentric and 
circular cases, dividing the two velocities 
gives the scaling for the 
measured increase in mass. The result is:
\begin{equation}\label{eq:mdot}
    \frac{\dot{M}-\dot{M}_0}{\dot{M}_0} = \frac{2 e}{q_t h_p}
\end{equation}
where the prefactor of $2$ is chosen empirically as an 
approximate value necessary to fit the data in Figure \ref{fig:mdot}.
This enhancement to the processed gas mass is then simply a 
consequence of the planet sweeping up more gas as it moves 
upon epicycles than it can on circular orbit.

\subsubsection{Midplane Streamlines}\label{sec:mp}
In section \ref{sec:rot} and 
particularly in Figure \ref{fig:rot}, it is evident that eccentric 
motion can enhance the rotational state of gas in the midplane. 
The eccentric motion also acts to enhance the radial motion of gas in
the midplane. 

\begin{figure*}
\gridline{
          \fig{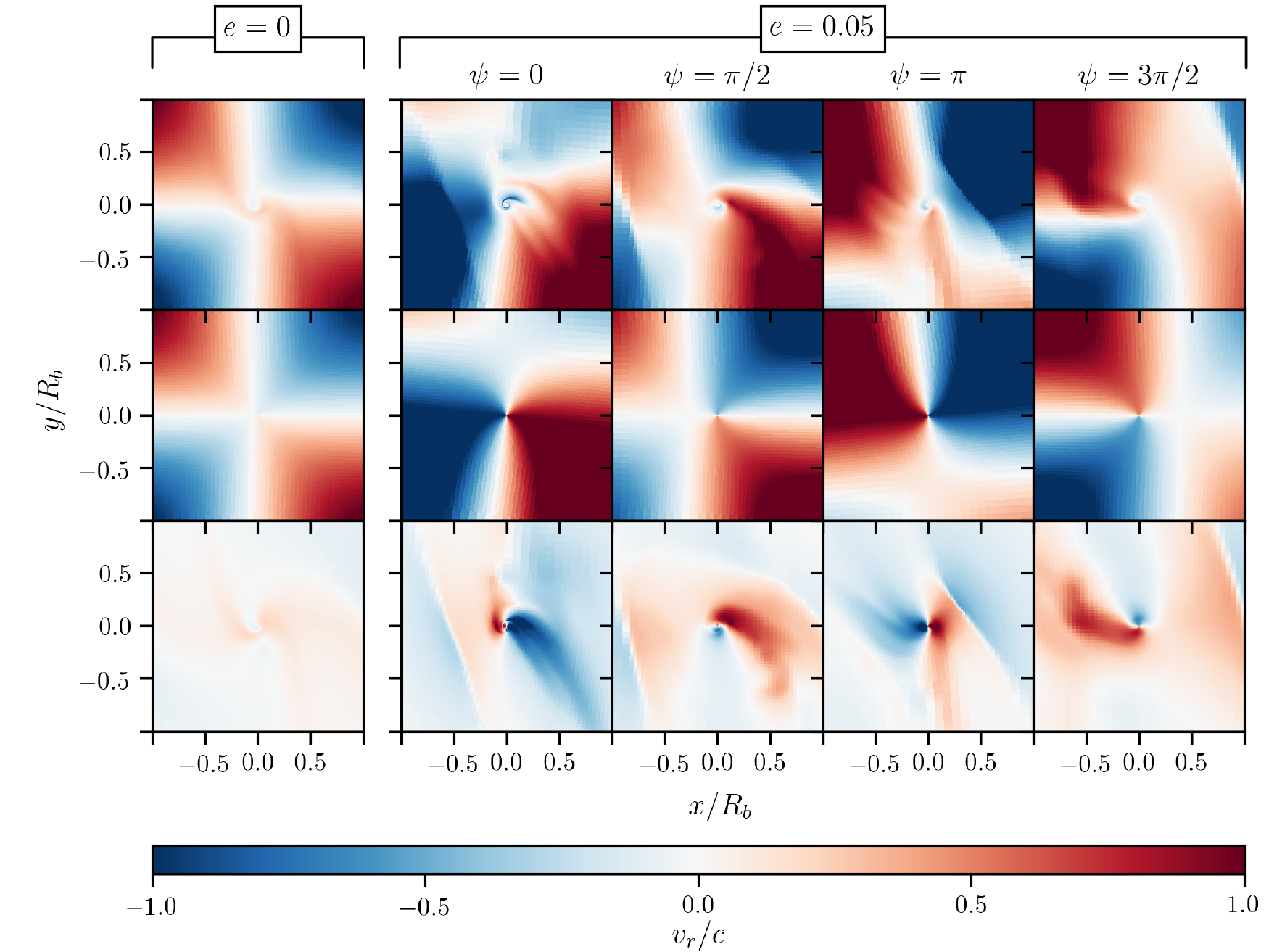}{0.98\textwidth}{}
          }
\caption{Midplane radial velocity $v_r$ in the planet frame for models 
\texttt{q25e0} and \texttt{q25e05}. 
The top row shows $v_r$ in our simulations. The middle row shows $v_r$ 
one expects arising purely from the disk's background shear 
(Equation \eqref{eq:v0}) and the 
epicyclic motion of the planet (Eqns. \eqref{eq:r}-\eqref{eq:phi}). 
The bottom row plots the residual of the first two rows, 
where positive (red) is planet-induced 
outflow and negative (blue) is planet-induced inflow. 
}
\label{fig:vr}
\end{figure*}

In the eccentric cases, the inflow is no 
longer restricted to $\phi=\pi/4$, $\phi=5\pi/4$ but is augmented 
by inflow from the planet plowing through the surrounding 
disk. This inflow changes orientation based on the phase of the orbit, 
traversing the full $2\pi$ in azimuth. 
These general characteristics of 
the midplane inflow/outflow can be 
seen in the top row of Figure \ref{fig:vr} where we plot the radial velocity 
in the midplane for several phases of an eccentric model along with the 
corresponding circular model. To decent approximation, the flow field 
can be considered as just the sum of the background shear and the 
epicyclic motion of the planet. This approximation is shown as the 
middle row in Figure \ref{fig:vr} and can be compared to the top row 
for an estimate of its accuracy. 
We also plot the residual in the bottom row, which represents the change to the flow field arising from 
the interaction with the planetary potential, 
which is typical for all eccentric models.
In all cases, the planet induces additional outflow in the direction of epicyclic motion and additional inflow in the opposite direction. These flows act to oppose gas motion supplied by epicyclic motion, as one might expect from the fact that the planet is creating a steady atmosphere around itself.
The magnitude of this induced inflow/outflow tends to increase 
with larger $q_t$ and larger $e$.

\begin{figure*}
\gridline{
          \fig{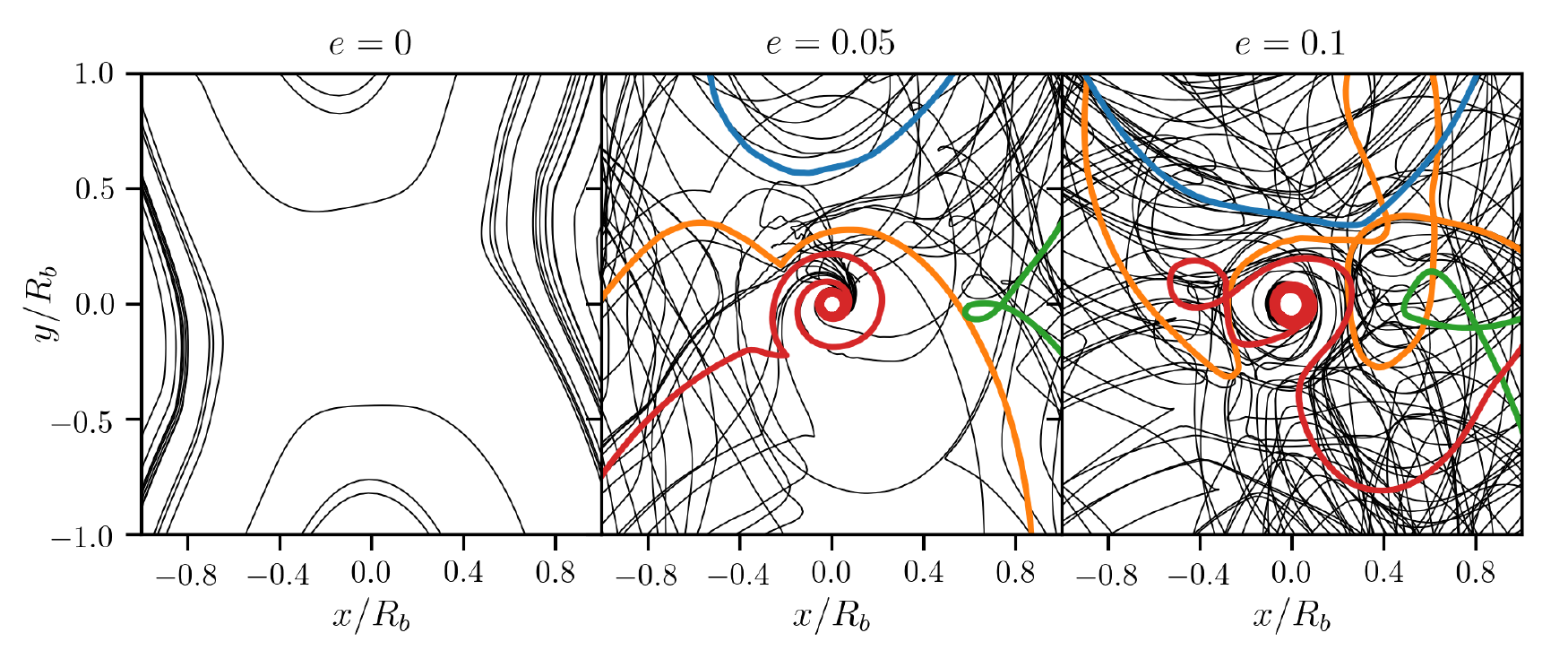}{\textwidth}{}
          }
\caption{Midplane pathlines for models \texttt{q1e0}, \texttt{q1e05}, and 
\texttt{q1e1}. Color is chosen to highlight particular pathlines 
(see text for details).
}
\label{fig:lines}
\end{figure*}

To capture the full midplane velocity field, we integrate 
tracer particle trajectories. We do this by post-processing models 
\texttt{q1e0}, \texttt{q1e05}, \texttt{q1e1}. 
$10,000$ particles are seeded randomly in the region $0.65<r_\ast<1.65$ and 
$-\pi/4<\phi_\ast<\pi/4$ but excluding a $1.5R_b$ around the planet. 
The trajectories are then integrated for 2 orbits 
using the full time-dependent flow field. We only use a first order method 
for this integration but with a timestep equal to the Courant condition, 
updating the flow field every timestep.
In Figure \ref{fig:lines} we plot a subset of these randomly seeded 
particle trajectories in the frame of the planet. For a more faithful 
representation of pathlines, the number of pathlines 
in each panel is kept proportional to number of particles that entered the 
frame. Due to our short integration time, some particle trajectories are 
"unfinished" -- the integration ends shortly after they enter the frame. 
We replace these few pathlines with other randomly selected pathlines 
that exhibit a more complete trajectory in Figure \ref{fig:lines}.

Upon integrating particle trajectories, 
it becomes clear that some fluid elements are able to make it 
through the time dependent inflow/outflow and travel all the way from 
$R_b$ to circulate near planet. 
Two such particle paths are highlighted in Figure 
\ref{fig:lines} in red. This behavior is not possible if the planet were on
a circular orbit
due to the robust outflowing nature of midplane gas near the planet.
In orange, we highlight a couple of particles that were 
able to make it fairly close to the planet but were ultimately swept up by 
outflow. Blue and green pathlines show particles that entered $R_b$ but 
are promptly diverted away either by outflow induced by the planet or 
outflow from the changing epicylic motion.

These particle trajectories illuminate several other key 
points of the eccentric midplane flow field. Most notable is just how 
different the geometry of the paths are between circular and eccentric 
models. Even for $e=0.05$, it becomes impossible to associate an 
eccentric pathline with one of the families of streamlines in the 
circular case (Figure \ref{fig:stream}). 
It is possible that the dearth of pathlines for $e=0.05$ 
just below the planet could be associated with the circular 
outflowing horseshoe and recycling streamlines however. 
We also draw attention to the fact that the 
$e=0.1$ contains a higher density of pathlines than $e=0.05$,
which in turn has a higher pathline density than the circular, 
$e=0$ case.
Because the same number of pathlines were randomly seeded in each 
model and we keep the number of displayed pathlines tied to the number 
of particles entering the frame, this higher density is partially a 
consequence of the mass flux enhancement discussed in \ref{sec:circ}. 
The model with higher $e$ sweeps up more material into $R_b$ as it 
undergoes its epicyclic motion.

\section{Implications}\label{sec:imp}

\subsection{Solid Accretion}\label{sec:peb}
Planets immersed in a Keplerian shear are expected to readily 
accrete solid pebble sized particles due to the combined effects 
of gas drag and the planetary potential. 
If the pebbles have settled into a disk thinner than $\sim R_b$,
we expect eccentricity to increase pebble accretion rate. 
\citet{LiuOrmel2018} show
that when the relative velocity between the planet and a pebble with
stopping time $t_s$ is set by eccentric motion ($v\propto e$), 
it leads to a smaller accretion cross section 
$b = \sqrt{GM_pt_s/v}\propto e^{-1/2}$ 
with increasing $e$. 
The net pebble mass flux however, is increased: 
$\dot{M}=\rho b v\propto e^{1/2}$. 
On the other hand, if the pebble layer is thick, 
the mass flux scales as $\dot{M}=\rho b^2 v$ so 
that eccentric motion causes no change to the accretion rate. 
For sufficiently high eccentricity 
($e/h_p\gtrsim (q_t/\tau_s)^{1/3}$), pebbles are not 
sufficiently coupled to the gas and pebbles with dimensionless 
stopping time $\tau_s$, enter the ballistic regime 
\citep{LiuOrmel2018}. 
In the ballistic regime, the cross section 
is set by the planet’s physical radius, leading to a severe 
suppression of the accretion rate. The result is an increase in the 
2D pebble accretion rate by a factor of $\sim 4$ for modest 
$e\sim0.05$ but not too large eccentricities.

This picture of eccentric pebble accretion does not include changes 
to flow field induced by the planetary potential however.
Using our high resolution 3D simulations, 
we have been able to test whether the eccentric flow field is sufficiently
determined by the eccentric motion or whether deviations arise from 
interactions with the planet. 

A first and obvious amendment to the 
estimates put forth by \citet{LiuOrmel2018} is motivated by the 
dramatic density profile surrounding the planet. 
In both circular and eccentric cases, 
the atmosphere density is orders of magnitude higher than 
the background disk. For a pebble in the Epstein drag regime, this 
density increase leads to an equivalent decrease in the stopping time
$t_s\propto \rho^{-1}$ and thereby a smaller accretion cross section 
$b\propto t_s^{1/2}\propto \rho^{-1/2}$. 
This skews the accreted pebble distribution toward larger pebble 
sizes. 
On the other hand, a smaller 
$t_s$ means a pebble is less likely to fall into ballistic regime, 
leading to a substantial increase in the accretion cross section 
for pebbles near the ballistic regime transition. 
With eccentric motion, we find evidence 
of mass increase interior to a bow shock which only adds to these 
effects. For pebbles experiencing Stokes drag, $t_s$ is 
independent of $\rho$, and these effects vanish.

For our steeply increasing density profile, pebbles able to 
make it close to the planet can have their stopping times 
reduced by 2-3 orders of magnitude. The fate of pebbles with 
sufficiently low $\tau_s$ to make them well-coupled to the 
gas becomes determined by the topology of the velocity field.
In the circular case there is no hope to be accreted — 
they are simply diverted away by the horseshoe flows 
\citep{Ormel2013,Kuwahara+2019} or recycling streamlines 
seen in our Figure \ref{fig:stream}. In the eccentric case, the 
flow topology changes drastically and judging from our red curves in 
Figure \ref{fig:lines}, there exist trajectories that advect particles 
significantly closer to the planet. If a pebble is advected by one 
of these trajectories, the possibility of accretion is increased, 
owing both to the increased gravity and the opportunity to 
circulate the planet for an extended period. We therefore expect 
the accretion rate of sufficiently low $\tau_s$ pebbles to be  
greater than the circular case. On the other hand, we find 
the planet induces outflow always oriented to oppose the epicyclic 
motion (Figure \ref{fig:vr}). Small inflowing 
particles can be slowed by this induced flow and then swept up 
by outflow as the planet changes the direction of its motion.
At least with tracer particles, we see a significant fraction 
penetrate into the Bondi radius, become stalled near the planet 
induced outflow, and then exit as the epicyclic motion changes 
large scale inflow to outflow. 
This would suggest some suppression for the accretion 
efficiency predicted by \citet{LiuOrmel2018} toward the small 
$t_s$ end. Because we still see some tracer particles making it close 
to the planet, the net result is still one of increased 
accretion relative to the circular case. 

For larger particles and those in the Stokes regime, the small scale 
flow topology and density profile become irrelevant. The accretion of 
these particles is instead determined by the large scale flows set by 
the planet’s eccentric motion. As we demonstrate in Section 
\ref{sec:circ}, these large scale fluxes are enhanced by a factor of 
up to $\sim 4$, simply due to the eccentric orbit sweeping up more gas 
as it moves through the disk. 
A similar enhancement may be applied to the pebble accretion rate.
Though they use different 
methods, \citet{LiuOrmel2018} also find a pebble accretion 
enhancement of up to $\sim 4$, likely because the underlying physics 
is similar. The details depend upon both pebble size and 
eccentricity, but our estimates show that the pebble accretion rate 
for eccentric planets should be a factor of a few greater than their 
circular counterparts.

\subsection{Gas Accretion}\label{sec:acc}
Traditional 1D models of core accretion treat the evolution as a 
hydrostatic growth sequence whose contraction is regulated by 
the envelope’s ability to cool. Our hydrodynamic simulations exhibit 
two departures from this classic picture: 1) A continuous recycling 
flux of unbound gas exterior to $r\gtrsim 0.1 R_b$ and 2) A robust 
bound CPD occurring within $r\lesssim 0.1R_b$. Eccentric motion tends 
to increase the former while leaving the latter unchanged.

To the first point -- circular orbit atmospheres both here and in 
previous studies \citep{Ormel+2015,Cimerman+2017,LambrechtsLega2017} 
exhibit significant recycling of unbound gas 
through at least some portion of their atmosphere. 
To quantify the rate of this recycling, \citet{Cimerman+2017} 
define a recycling time as the mass of the 
envelope interior to some radius divided by the inflowing mass flux: 
$t_{\rm recycling} = M(<r)/\dot{M}_{\rm in}(r)$. 
When this recycling time is less than the cooling time for the gas, 
gas cannot cool efficiently. In our eccentric models, the recycling 
time can be made significantly smaller by the mass flux enhancement
discussed in Section \ref{sec:circ}. Because the mass of 
an eccentric model is roughly the same as its circular equivalent, 
the eccentric $t_{\rm recycling}$ can be obtained simply by 
dividing the circular recycling time by our Equation \eqref{eq:mdot}. 
If the mass flux enhancement
varies as a function of radius then this should also be taken 
into account. However we find that eccentric mass flux enhancement 
also tends to be fairly constant over a large range of radius. 
The extent of the recycling region and its importance varies 
depending on the equation of state and how one defines gas as 
‘bound’. In our case, recycling is best defined as occurring 
exterior to the CPD, $r\gtrsim 0.1R_b$. In any case, eccentric 
motion should only serve to hasten the recycling of gas irrespective 
of one’s precise definition of the recycling region.

To the second point -- interior to $0.1R_b$, 
the flow is dictated by the CPD. 
Similar to \citet{Fung+2019}, the gas in our eccentric CPDs is 
bound in a kinematic sense — orbit-averaged midplane radial 
velocities are directed towards the planet within the CPD and 
away from the planet exterior to the CPD. 
In this case, the eventual accretion of gas onto the planet would 
be regulated by details of the CPD with the recycled outer envelope 
acting as an outer boundary condition. In this sense, eccentricity 
only matters for setting the boundary condition on the CPD. 

If eccentricity were able to modify the rotation curve of the CPD, 
it would also directly modify the accretion rate, but as we show 
in Section \ref{sec:rot}, the CPD rotation curve remains insensitive 
to changes in $e$. This is at least partially a consequence of 
our choice of an isothermal equation of state.
Under a different equation of state, rotational velocities tend to be 
lower than in our isothermal models \citep{Fung+2019} and often fail 
to form CPDs until well above our mass range \citep{Schulik+2020}. 
In those cases, where gas is rotating more slowly, 
the rotation curve could be more dramatically modified 
by eccentric motion (akin to our \texttt{hs} models). 
If the rotational velocities were increased/decreased (as in our 
subsonic/supersonic case), the accretion rate could 
be suppressed/enhanced. But because the 3D kinematics and 
thermodynamics are inextricably linked, simulations including the 
proper prescriptions for cooling are required to address these 
questions.

\subsection{Eccentricity Damping}\label{sec:damp}
With some idea of the flow field around eccentric planets, we call 
into question the often made assumption that sub-Jovian planets 
should have their eccentricity damped. As we have 
seen, our model planets are able to significantly perturb the 
flow field in non-linear and non-trivial ways. Though one can 
attempt to estimate non-linear torques in the method of
\citet{Ward1991} by including the angular momentum exchanged between 
the planet and horseshoe streamlines, our simulations demonstrate 
their inadequacy. Figure \ref{fig:lines} shows that the notion of 
horseshoes, separatrices, families of streamlines, all fail to 
apply to even the modest eccentricity of $e=0.05$. This suggests a 
need for full 3D simulations studying the eccentricity evolution of 
protoplanets. To our knowledge however, only one such study exists --  
\citet{BitschKley2010}. Though their work finds a general pattern of 
eccentricity damping, their models suffer from a resolution of $\sim 5$ 
cells per $R_b$ and the excision of a significant fraction of $R_b$. 
Though this excision was a sensible effort to disregard contributions 
from gravitationally bound material, our models show that most of 
this material is actually unbound. Judging from the complicated flow 
pattern of our Figure \ref{fig:lines}, it seems that $\sim 5$ cells 
per $R_b$ is insufficient to resolve the torque contribution near 
the planet. With no other such simulations performed in the 
past decade, we find insufficient evidence to conclude that 
sub-Jovian protoplanets should have their eccentricities damped. 
We are working on calculating the torques for our own simulations
in a self-consistent way, in order to determine the resulting orbital 
evolution.

\section{Summary}
We have carried out a set of global 3D isothermal 
simulations of eccentric planets with $e\leq 0.1$ 
and thermal mass $0.25 < q_t < 2.0$ corresponding 
to $10 M_\oplus < M_p < 80 M_\oplus$ around a $1M_\odot$ star.
Our main findings are:
\begin{enumerate}
    \item Supersonic eccentric motion leads to the formation of a bow 
    shock. This increases the overall mass interior to the shock (Section \ref{sec:den}).
    \item The structure and rotation of the CPD is resilient, remaining
    kinematically bound and similar to the circular orbit models even with a moderate eccentricity (Section \ref{sec:rot}).
    \item Exterior to the CPD, subsonic eccentric motion leads to 
    increased prograde rotation. This occurs for subthermal 
    models with sufficiently large yet still subsonic $e$.
    As the motion becomes supersonic, the rotation in subthermal models becomes increasingly retrograde (Section \ref{sec:rot}).
    \item Flow geometry, now dependent on orbital phase, 
    is altered from the established circular picture. Instead of a 
    pattern of polar inflow/midplane outflow, it can 
    vary wildly. In some cases, the amount of midplane inflow becomes 
    greater than the polar inflow (Section \ref{sec:circ}).
    \item Flow magnitudes and fluxes are amplified with 
    velocity set by the eccentric 
    motion. The flux of material processed through the Bondi 
    radius is enhanced by a factor $\sim 2 e/(q_t h_p)$. 
    Recycling times exterior to the CPD are 
    decreased by this same factor (Section \ref{sec:mp}).
    \item The increased flux 
    can increase 
    the pebble accretion rate 
    up to several times the circular rate. Planetary perturbations to 
    flow field can suppress this accretion rate for the smallest 
    solids, but the net result is still one of increased pebble 
    accretion (Section \ref{sec:peb}).
    \item Because the CPD is unaltered by eccentric motion, the rate 
    of gas accretion onto eccentric planets should be comparable to 
    the circular case (Section \ref{sec:acc}).
    
\end{enumerate}
Though we make some extrapolations to lower mass protoplanets, 
a global framework limits the ability 
to explore the subthermal regime more thoroughly. Future models, employing 
the local approximation, will be able to test whether the conclusions made 
here still hold when extrapolated to lower $q_t$. Future models will also 
implement a more sophisticated treatment of the thermodynamics. This 
will undoubtedly change the nature of the bow shock interaction 
among other things. Additionally, though we make some qualitative 
predictions about the accretion of solids, quantitative verification 
requires integrating solid trajectories with the included 
effects of gas drag. 
With such a time-dependent flow, it becomes necessary to integrate the 
particles in conjunction with the hydrodynamics. With the development of 
a dust module in the \textsc{Athena++} code, we will be able to more 
adequately address this question of solid accretion onto eccentric 
protoplanets. Regardless, this paper serves a departure from the oft 
assumed circular orbit and demonstrates that small to moderate 
eccentricities can alter established characteristics of planet formation.

\bibliography{planet_ecc}{}

\begin{thebibliography}{}
\expandafter\ifx\csname natexlab\endcsname\relax\def\natexlab#1{#1}\fi
\providecommand{\url}[1]{\href{#1}{#1}}
\providecommand{\dodoi}[1]{doi:~\href{http://doi.org/#1}{\nolinkurl{#1}}}
\providecommand{\doeprint}[1]{\href{http://ascl.net/#1}{\nolinkurl{http://ascl.net/#1}}}
\providecommand{\doarXiv}[1]{\href{https://arxiv.org/abs/#1}{\nolinkurl{https://arxiv.org/abs/#1}}}

\bibitem[{{Artymowicz}(1993)}]{Artymowicz1993}
{Artymowicz}, P. 1993, \apj, 419, 166, \dodoi{10.1086/173470}

\bibitem[{{Ayliffe} \& {Bate}(2009)}]{AyliffeBate2009}
{Ayliffe}, B.~A., \& {Bate}, M.~R. 2009, \mnras, 397, 657,
  \dodoi{10.1111/j.1365-2966.2009.15002.x}

\bibitem[{{Ayliffe} \& {Bate}(2012)}]{AyliffeBate2012}
---. 2012, \mnras, 427, 2597, \dodoi{10.1111/j.1365-2966.2012.21979.x}

\bibitem[{{Batygin} \& {Morbidelli}(2020)}]{BatyginMorbidelli2020}
{Batygin}, K., \& {Morbidelli}, A. 2020, \apj, 894, 143,
  \dodoi{10.3847/1538-4357/ab8937}

\bibitem[{{B{\'e}thune} \& {Rafikov}(2019)}]{BethuneRafikov2019}
{B{\'e}thune}, W., \& {Rafikov}, R.~R. 2019, \mnras, 488, 2365,
  \dodoi{10.1093/mnras/stz1870}

\bibitem[{{Bitsch} \& {Kley}(2010)}]{BitschKley2010}
{Bitsch}, B., \& {Kley}, W. 2010, \aap, 523, A30,
  \dodoi{10.1051/0004-6361/201014414}

\bibitem[{{Chrenko} {et~al.}(2017){Chrenko}, {Bro{\v{z}}}, \&
  {Lambrechts}}]{Chrenko+2017}
{Chrenko}, O., {Bro{\v{z}}}, M., \& {Lambrechts}, M. 2017, \aap, 606, A114,
  \dodoi{10.1051/0004-6361/201731033}

\bibitem[{{Cimerman} {et~al.}(2017){Cimerman}, {Kuiper}, \&
  {Ormel}}]{Cimerman+2017}
{Cimerman}, N.~P., {Kuiper}, R., \& {Ormel}, C.~W. 2017, \mnras, 471, 4662,
  \dodoi{10.1093/mnras/stx1924}

\bibitem[{{D'Angelo} \& {Bodenheimer}(2013)}]{DangeloBodenheimer2013}
{D'Angelo}, G., \& {Bodenheimer}, P. 2013, \apj, 778, 77,
  \dodoi{10.1088/0004-637X/778/1/77}

\bibitem[{{D'Angelo} {et~al.}(2006){D'Angelo}, {Lubow}, \&
  {Bate}}]{Dangelo+2006}
{D'Angelo}, G., {Lubow}, S.~H., \& {Bate}, M.~R. 2006, \apj, 652, 1698,
  \dodoi{10.1086/508451}

\bibitem[{{Duffell} \& {Chiang}(2015)}]{DuffellChiang2015}
{Duffell}, P.~C., \& {Chiang}, E. 2015, \apj, 812, 94,
  \dodoi{10.1088/0004-637X/812/2/94}

\bibitem[{{Dunhill} {et~al.}(2013){Dunhill}, {Alexander}, \&
  {Armitage}}]{Dunhill+2013}
{Dunhill}, A.~C., {Alexander}, R.~D., \& {Armitage}, P.~J. 2013, \mnras, 428,
  3072, \dodoi{10.1093/mnras/sts254}

\bibitem[{{Eklund} \& {Masset}(2017)}]{EklundMasset2017}
{Eklund}, H., \& {Masset}, F.~S. 2017, \mnras, 469, 206,
  \dodoi{10.1093/mnras/stx856}

\bibitem[{{Fendyke} \& {Nelson}(2014)}]{FendykeNelson2014}
{Fendyke}, S.~M., \& {Nelson}, R.~P. 2014, \mnras, 437, 96,
  \dodoi{10.1093/mnras/stt1867}

\bibitem[{{Fung} {et~al.}(2015){Fung}, {Artymowicz}, \& {Wu}}]{Fung+2015}
{Fung}, J., {Artymowicz}, P., \& {Wu}, Y. 2015, \apj, 811, 101,
  \dodoi{10.1088/0004-637X/811/2/101}

\bibitem[{{Fung} {et~al.}(2017){Fung}, {Masset}, {Lega}, \&
  {Velasco}}]{Fung+2017}
{Fung}, J., {Masset}, F., {Lega}, E., \& {Velasco}, D. 2017, \aj, 153, 124,
  \dodoi{10.3847/1538-3881/153/3/124}

\bibitem[{{Fung} {et~al.}(2014){Fung}, {Shi}, \& {Chiang}}]{Fung+2014}
{Fung}, J., {Shi}, J.-M., \& {Chiang}, E. 2014, \apj, 782, 88,
  \dodoi{10.1088/0004-637X/782/2/88}

\bibitem[{{Fung} {et~al.}(2019){Fung}, {Zhu}, \& {Chiang}}]{Fung+2019}
{Fung}, J., {Zhu}, Z., \& {Chiang}, E. 2019, \apj, 887, 152,
  \dodoi{10.3847/1538-4357/ab53da}

\bibitem[{{Goldreich} \& {Sari}(2003)}]{GoldreichSari2003}
{Goldreich}, P., \& {Sari}, R. 2003, \apj, 585, 1024, \dodoi{10.1086/346202}

\bibitem[{{Kley}(1998)}]{Kley1998}
{Kley}, W. 1998, \aap, 338, L37.
\newblock \doarXiv{astro-ph/9808351}

\bibitem[{{Kurokawa} \& {Tanigawa}(2018)}]{Kurokawa+2018}
{Kurokawa}, H., \& {Tanigawa}, T. 2018, \mnras, 479, 635,
  \dodoi{10.1093/mnras/sty1498}

\bibitem[{{Kuwahara} \& {Kurokawa}(2020)}]{KuwaharaKurokawa2020}
{Kuwahara}, A., \& {Kurokawa}, H. 2020, \aap, 633, A81,
  \dodoi{10.1051/0004-6361/201936842}

\bibitem[{{Kuwahara} {et~al.}(2019){Kuwahara}, {Kurokawa}, \&
  {Ida}}]{Kuwahara+2019}
{Kuwahara}, A., {Kurokawa}, H., \& {Ida}, S. 2019, \aap, 623, A179,
  \dodoi{10.1051/0004-6361/201833997}

\bibitem[{{Lambrechts} \& {Lega}(2017)}]{LambrechtsLega2017}
{Lambrechts}, M., \& {Lega}, E. 2017, \aap, 606, A146,
  \dodoi{10.1051/0004-6361/201731014}

\bibitem[{{Liu} \& {Ormel}(2018)}]{LiuOrmel2018}
{Liu}, B., \& {Ormel}, C.~W. 2018, \aap, 615, A138,
  \dodoi{10.1051/0004-6361/201732307}

\bibitem[{{Machida}(2009)}]{Machida2009}
{Machida}, M.~N. 2009, \mnras, 392, 514,
  \dodoi{10.1111/j.1365-2966.2008.14090.x}

\bibitem[{{Machida} {et~al.}(2008){Machida}, {Kokubo}, {Inutsuka}, \&
  {Matsumoto}}]{Machida+2008}
{Machida}, M.~N., {Kokubo}, E., {Inutsuka}, S.-i., \& {Matsumoto}, T. 2008,
  \apj, 685, 1220, \dodoi{10.1086/590421}

\bibitem[{{Mai} {et~al.}(2020){Mai}, {Desch}, {Kuiper}, {Marleau}, \&
  {Dullemond}}]{Mai+2020}
{Mai}, C., {Desch}, S.~J., {Kuiper}, R., {Marleau}, G.-D., \& {Dullemond}, C.
  2020, \apj, 899, 54, \dodoi{10.3847/1538-4357/aba4a8}

\bibitem[{{Mignone} {et~al.}(2007){Mignone}, {Bodo}, {Massaglia}, {Matsakos},
  {Tesileanu}, {Zanni}, \& {Ferrari}}]{Mignone+2007}
{Mignone}, A., {Bodo}, G., {Massaglia}, S., {et~al.} 2007, \apjs, 170, 228,
  \dodoi{10.1086/513316}

\bibitem[{{Mizuno}(1980)}]{Mizuno1980}
{Mizuno}, H. 1980, Progress of Theoretical Physics, 64, 544,
  \dodoi{10.1143/PTP.64.544}

\bibitem[{{Muley} {et~al.}(2019){Muley}, {Fung}, \& {van der
  Marel}}]{Muley+2019}
{Muley}, D., {Fung}, J., \& {van der Marel}, N. 2019, \apjl, 879, L2,
  \dodoi{10.3847/2041-8213/ab24d0}

\bibitem[{{Ogilvie} \& {Lubow}(2002)}]{OgilvieLubow2002}
{Ogilvie}, G.~I., \& {Lubow}, S.~H. 2002, \mnras, 330, 950,
  \dodoi{10.1046/j.1365-8711.2002.05148.x}

\bibitem[{{Ormel}(2013)}]{Ormel2013}
{Ormel}, C.~W. 2013, \mnras, 428, 3526, \dodoi{10.1093/mnras/sts289}

\bibitem[{{Ormel} \& {Liu}(2018)}]{OrmelLiu2018}
{Ormel}, C.~W., \& {Liu}, B. 2018, \aap, 615, A178,
  \dodoi{10.1051/0004-6361/201732562}

\bibitem[{{Ormel} {et~al.}(2015){Ormel}, {Shi}, \& {Kuiper}}]{Ormel+2015}
{Ormel}, C.~W., {Shi}, J.-M., \& {Kuiper}, R. 2015, \mnras, 447, 3512,
  \dodoi{10.1093/mnras/stu2704}

\bibitem[{{Papaloizou} \& {Larwood}(2000)}]{PapaloizouLarwood2000}
{Papaloizou}, J.~C.~B., \& {Larwood}, J.~D. 2000, \mnras, 315, 823,
  \dodoi{10.1046/j.1365-8711.2000.03466.x}

\bibitem[{{Pollack} {et~al.}(1996){Pollack}, {Hubickyj}, {Bodenheimer},
  {Lissauer}, {Podolak}, \& {Greenzweig}}]{Pollack+1996}
{Pollack}, J.~B., {Hubickyj}, O., {Bodenheimer}, P., {et~al.} 1996, \icarus,
  124, 62, \dodoi{10.1006/icar.1996.0190}

\bibitem[{{Schulik} {et~al.}(2019){Schulik}, {Johansen}, {Bitsch}, \&
  {Lega}}]{Schulik+2019}
{Schulik}, M., {Johansen}, A., {Bitsch}, B., \& {Lega}, E. 2019, \aap, 632,
  A118, \dodoi{10.1051/0004-6361/201935473}

\bibitem[{{Schulik} {et~al.}(2020){Schulik}, {Johansen}, {Bitsch}, {Lega}, \&
  {Lambrechts}}]{Schulik+2020}
{Schulik}, M., {Johansen}, A., {Bitsch}, B., {Lega}, E., \& {Lambrechts}, M.
  2020, arXiv e-prints, arXiv:2003.13398.
\newblock \doarXiv{2003.13398}

\bibitem[{{Stone} {et~al.}(2020){Stone}, {Tomida}, {White}, \&
  {Felker}}]{Stone+2020}
{Stone}, J.~M., {Tomida}, K., {White}, C.~J., \& {Felker}, K.~G. 2020, \apjs,
  249, 4, \dodoi{10.3847/1538-4365/ab929b}

\bibitem[{{Szul{\'a}gyi}(2017)}]{Szulagyi2017}
{Szul{\'a}gyi}, J. 2017, \apj, 842, 103, \dodoi{10.3847/1538-4357/aa7515}

\bibitem[{{Szul{\'a}gyi} {et~al.}(2016){Szul{\'a}gyi}, {Masset}, {Lega},
  {Crida}, {Morbidelli}, \& {Guillot}}]{Szulagyi+2016}
{Szul{\'a}gyi}, J., {Masset}, F., {Lega}, E., {et~al.} 2016, \mnras, 460, 2853,
  \dodoi{10.1093/mnras/stw1160}

\bibitem[{{Tanigawa} {et~al.}(2012){Tanigawa}, {Ohtsuki}, \&
  {Machida}}]{Tanigawa+2012}
{Tanigawa}, T., {Ohtsuki}, K., \& {Machida}, M.~N. 2012, \apj, 747, 47,
  \dodoi{10.1088/0004-637X/747/1/47}

\bibitem[{{Wang} {et~al.}(2014){Wang}, {Bu}, {Shang}, \& {Gu}}]{Wang+2014}
{Wang}, H.-H., {Bu}, D., {Shang}, H., \& {Gu}, P.-G. 2014, \apj, 790, 32,
  \dodoi{10.1088/0004-637X/790/1/32}

\bibitem[{{Ward}(1991)}]{Ward1991}
{Ward}, W.~R. 1991, in Lunar and Planetary Science Conference, Vol.~22, Lunar
  and Planetary Science Conference, 1463

\end{thebibliography}
\bibliographystyle{aasjournal}

\end{document}